\documentclass{aa}
\usepackage{txfonts}
\usepackage{graphicx}
\usepackage{mathtools}
\usepackage{ulem}

\makeatletter
\renewcommand*\aa@pageof{, page \thepage{} of \pageref*{LastPage}}
\makeatother

\begin{document}

\title{The influence of planetesimal fragmentation on planet formation}
\author{Nicolas Kaufmann\inst{1}, Yann Alibert\inst{1}}

\institute{Physikalisches Institut, Universität Bern, Gesselschaftsstrasse 6, 3012 Bern, Switzerland}

\date{Received 13/01/2023 / Accepted 26/05/2023}

\abstract{The effects of planetesimal fragmentation on planet formation has been studied by various models on single embryos {therefore neglecting concurrent effects} mostly in the outer disk. They show that planetesimal fragmentation can both hinder or aid planet formation due to the introduction of competing effects, namely speeding up accretion and depleting the feeding zone of forming planets.}
{We investigate the {influence} of the collisional fragmentation of planetesimals on the planet formation process {using} a population synthesis approach. We aim to investigate its effects {for a large set of initial conditions} and also explore {the consequences on} the formation of multiple embryos in the same disk.}
{We run global planet formation simulations {including} fragmentation, drift and an improved ice line description. To do this we {use} a fragmentation model in our code. The initial conditions for the simulations that are informed by observations are varied to generate synthetic exoplanet populations.} 
{Our synthetic populations show that depending on the typical size of solids generated in collisions, fragmentation {in tandem with the radial drift} can either enhance or hinder planet formation. For larger fragments we see increased accretion throughout the populations especially beyond the ice line. However, {the shorter drift timescale of} smaller fragments, {due to their stronger coupling to the gas}, can hinder the formation process. Furthermore, beyond the ice line fragmentation promotes late growth {when the damping by gas drag fades}}
{Fragmentation significantly affects the planet formation process in various ways for all types of planet and warrants further investigation.}


\titlerunning{the influence of planetesimal fragmentation
on synthetic planet populations}
  
\maketitle

\section{Introduction}
In the classical core accretion scenario the terrestrial planets and cores of giant planets are formed in multiple steps. The first step is the formation of planetesimals from dust {for which} multiple pathways {are proposed}. The two main proposed mechanisms are coagulation and the gravitational collapse via streaming instability \citep{Okuzumi_2012,Kobayashi_2021,Yang_2017}. The first process describes the formation of planetesimals via sticking of smaller dust grains in collisions. In the second process, streaming instability forms planetesimals through the gravitational collapse of chondrule-sized objects {\citep{Johansen_2007,Johansen_2014,Schaefer_2017}}.\\
From these planetesimals planetary embryos form via runaway growth as their accretion rate speeds up with increasing mass so they separate themselves from the {rest of the} planetesimals \citep{Ormel_2010}. As the embryos grow they start stirring up the remaining planetesimals which reduces the embryos accretion rate and allows other embryos (oligarchs) in neighbouring regions to catch up {in mass} \citep{Ida_1993}. The size distribution of the remaining planetesimals is quite important for the further growth of the embryos as the accretion rates depend strongly on their size and dynamical state \citep{Fortier_2012,Chambers_2006,Guilera_2010}. Furthermore, the size distribution of the planetesimals continues to evolve through their mutual collisions which can lead to coagulation or fragmentation of planetesimals depending on their relative speeds \citep{Kenyon_2004}. The relative speed between planetesimals is caused by their gravitational interaction with the embryos and other planetesimals and is influenced by the interactions with the gas disk. For higher relative speeds this results in the generation of smaller fragments that get removed from the disk via drift due to the sub-Keplerian headwind of the gas disk. In some {models} the initial size of the planetesimals is estimated to be $\approx 100 km$ \citep{Morbidelli_2009}, but the typical size of accreted solids is still poorly constrained \citep{Helled_2021}. In addition the {outcome of collisions} are also rather uncertain \citep{Kobayashi_2009} which makes the further study of the collisional evolution important to improve our understanding of these processes and their imprint on planet formation as a whole.

There have been many studies of the collisional evolution and fragmentation of planetesimals \citep{Kobayashi_2018,Chambers_2008,Guilera_2014,San_Sebastian_2019,Chambers_2014,Inaba_2003}. These studies show that planetesimal fragmentation can both inhibit and/or enhance the formation of planets as it introduces competing effects. This especially affects the growth of {the core of} gas giants as their greater mass increases the random velocities of planetesimals more strongly and therefore shortening collisonal timescales and the typical size of solids \citep{Guilera_2014}. These smaller solids are easier to accrete but also lead to the removal of accretable material caused by their faster drift speeds. 

Previous studies have been focused on single planet systems and often other parts of the planet formation process {have to be simplified for example} migration and the calculation of the internal structure of the forming planets. Furthermore, there have been no studies on the impact of fragmentation in a population synthesis approach which is valuable as it allows to probe a larger part of the parameter space of planet formation. The study of multiple discrete interacting embryos has also been neglected in previous studies but {its consideration is important as} it allows the investigation of features that may arise due to the presence of multiple planets.

In this work we {therefore} present a population level investigation of the influence of the size evolution of planetesimals on planet formation, more specifically the influence of the fragmentation of planetesimals. This is done by implementing a fragmentation model into the Bern {model} \citep{Alibert_2004,Alibert_2005,Mordasini_2009,Mordasini_2012,Alibert_2013,NGPPS1} and {by improving} the treatment of the solid disk. {This is then used to run many global planet formation simulations varying the initial conditions in a population synthesis approach. We consider the formation of single embryos or multiple embryos simultaneously. This lets us create synthetic planet populations that show how collisions between the solids in the disks influence the formation of various types of planets. To do this we run the planet formation model for different model choices and parameters of our fragmentation model to explore its effects in different regimes.}

In Sect. \ref{Model} we give an overview of the Bern model. We discuss the basic features of the code and the description of the solid disk due to its importance to the investigated collisional processes i.e. fragmentation. Furthermore we describe the newly implemented features that build on previous iterations {of the model} including drift and the calculation of the dynamical state. Then the addition of the fragmentation model that is introduced to the code {is} discussed. Finally The population synthesis approach will {is presented}. In section \ref{test},  {we test our calculations against other} formation models that include fragmentation to ensure its validity. Then we use the model to generate synthetic populations of planets in Sect. \ref{pop} to investigate the influence of fragmentation on the planet formation process. Finally we discuss and summarise our results and list our conclusions which is shown in Sect. \ref{res}. 

\section{Planet formation model}
\label{Model}
The numerical code used to model the formation of planets (the Bern model) is an adaption of the one described in \cite{NGPPS1}. The Bern model includes the formation and long term evolution of planets in two stages. The planet formation is first tracked for a fixed time interval (20 Myr) and is derived from the works of \citep{Alibert_2004,Alibert_2005,Mordasini_2009,Alibert_2013,NGPPS1}. {The gas disk is described as a viscous accretion disk. The turbulent viscosity is parameterised with the constant $\alpha$ parameter \cite{Shakura_1973}}. {The formation phase} additionally considers the interplay between the planet and the disk: migration, gas and solid accretion. The gravitational interactions between planetary embryos are integrated with the mercury integrator \citep{Chambers_1999}. The gas accretion is calculated by solving the structure equations. When the planet becomes massive enough, runaway gas accretion occurs, the envelop contracts and accretion is limited by the supply of the gas disk.
{we include core growth by accreting planetesimals in the oligarchic regime. For multiple planets, we consider up to one hundred initial embryos of moon mass ($10^{-2}M_\oplus$) and a mono-disperse swarm of planetesimals (when fragmentation is not considered) with a single size (typically $300m$)}. 

After the formation phase we continue to evolve the planets individually to $10$ Gyr as described in \citep{Mordasini_2012}. This includes the solving of the internal structure equations, atmospheric escape \citep{Jin_2014} and tidal migration \citep{Llambay_2011}. 
\subsection{solid disk}
\label{sec:soliddisk}
The planetesimals are not represented individually {but rather they are} described on a grid by a few key quantities, namely their surface density ($\Sigma$) their mean root squared eccentricity ($e$) and inclination ($i$), their typical size ($s$) and their bulk density ($\rho$) along with the ice fraction of the planetesimals. The introduction of fragmentation will add objects of different typical sizes {(fragments)} to the solid disk. To treat this we have separate grids (with individual $\Sigma$ {and other quantities}) for the fragments and planetesimals which we will also refer to as different swarms of solids {in the rest of the paper.}

The eccentricities and inclinations are assumed to be Rayleigh distributed which is motivated by N-body simulations \citep{Salo_1985}. Since we assume an azimuthally symmetrical disk we can describe the dynamical state i.e. their random velocities with just the mean root squared eccentricity and inclination of the swarm. The random velocities of the planetesimals, described by $e$ and $i$, are increased due to the gravitational interactions between the planetesimals {among} themselves and with the embryos and are dampened by the gas drag. The stirring by density fluctuations in the gas is also considered. To calculate the dynamical state we follow the approach of \citep{Fortier_2012} and solve the evolution equations for $e$ and $i$ at each time step given by,
\begin{align}
    \dot{e}^2 &= \dot{e}^2\big|_{drag} + \dot{e}^2\big|_{pp} + \dot{e}^2\big|_{stirr} + \dot{e}^2\big|_{DF} \\
    \dot{i}^2 &= \dot{i}^2\big|_{drag} + \dot{i}^2\big|_{pp} + \dot{i}^2\big|_{stirr} + \dot{i}^2\big|_{DF}, 
    \label{eq:dynamical_evolution}
\end{align}
where the contributions on the right hand side arise due to the gas drag, the planetesimal-planetesimal interaction, the stirring by the embryos and stirring by density fluctuations in that order.
The gas drag depends on the relative velocity between the planetesimals and the gas. We consider three regimes: Epstein, Stokes and quadratic \citep{Rafikov_2004}. The regimes are separated by the {size of the planetesimals and the} relative velocity between the planetesimals and the gas: $v_{rel} = \sqrt{\eta^2 + 5/8 e^2 + 1/2 i^2}$ where $\eta = -\frac{1}{2\Omega r \rho_{mid}}\frac{\partial p}{\partial r}$ is the deviation of the gas from Keplerian speed. The Epstein drag is considered when the planetesimals are roughly smaller then the mean free path of the gas i.e. $s< 1.5\times\lambda = (n_{H2} \sigma_{H2})^{-1}$ where $n$ is the number density of hydrogen which is the main component of the gas disk and $\sigma$ its cross-section. Otherwise the distinction between the two remaining cases is made with the Reynolds number $Re_{mol} = v_{rel}s/\nu_{mol}$ where  and $\nu_{mol} = \lambda c_s/3$ is the molecular viscosity and $c_s$ the sound speed. For a Reynolds number above $27$ the quadratic regime is considered and below the Stokes regime. The drag expressions for the Epstein regime are
\begin{align}
    \dot{e^2}\big|_{Edrag} &= -e^2 \frac{c_s\rho_{mid}}{\rho s} &\text{for}\:s<1.5\lambda\\
    \dot{i^2}\big|_{Edrag} &= -i^2/2 \frac{c_s\rho_{mid}}{\rho s}. 
\end{align}
The Formulas for the Stokes regime are
\begin{align}
    \dot{e^2}\big|_{Sdrag} &= -\frac{3e^2}{2} \frac{\lambda\rho_{mid}}{\rho s^2} &\text{for}\:s>1.5\lambda\\
    \dot{i^2}\big|_{Sdrag} &= -\frac{3i^2}{4} \frac{\lambda \rho_{mid}}{\rho s^2}&\text{and}\:Re_{mol}<27, 
\end{align}
and in the quadratic regime they are
\begin{align}
    \dot{e^2}\big|_{Qdrag} &= -2e^2 \frac{v_{rel}\rho_{mid}}{6\rho s} &\text{for}\:s>1.5\lambda\\
    \dot{i^2}\big|_{Qdrag} &= -i^2\frac{v_{rel} \rho_{mid}}{6\rho s}&\text{and}\:Re_{mol}>27.
\end{align}
The second term in the dynamical evolution stems from the planetesimal-planetesimal interaction and follows the description of \cite{Ohtsuki_2002},
\begin{align}
\label{eq:pp_stirr_e}
    \dot{e^2}\big|_{pp} = &a_0 \Omega \sum_j \big [ N_j \frac{h_{ij}^4 m_j}{( m_i+m_j)^2}* \big \{ m_j P_{VS}(\Tilde{e_i},\Tilde{i_i},\Tilde{e_j},\Tilde{i_j}) \nonumber \\&+ \frac{m_j e_j^2 - m_i e_i^2}{e_i^2 + e_j^2} P_{DF}  (\Tilde{e_i},\Tilde{i_i},\Tilde{e_j},\Tilde{i_j}) \big \} \big ] \\
    \label{eq:pp_stirr_i}
    \dot{i^2}\big|_{pp} = &a_0 \Omega \sum_{j} \big [\ N_j \frac{h_{ij}^4 m_j}{(\ m_i+m_j)\ ^2}* \big \{ m_j Q_{VS}(\Tilde{e_i},\Tilde{i_i},\Tilde{e_j},\Tilde{i_j}) \nonumber  \\
    &+ \frac{m_j i_j^2 - m_i i_i^2}{i_i^2 + i_j^2} Q_{DF} (\Tilde{e_i},\Tilde{i_i},\Tilde{e_j},\Tilde{i_j}) \big \} \big ]\ .
\end{align}

$a_0$ is the semi major axis, $\Omega$ is Kepler angular speed and $i$ and $j$ refer to the different swarms for example when we consider planetesimals {and fragments}. The viscous stirring is the first term in the {curly brackets in Eqs. \ref{eq:pp_stirr_e} and \ref{eq:pp_stirr_i}}. The dynamical friction term is second contribution which vanishes for a single size of planetesimals {i.e. when no fragments are present}. The functions $Q_{VS}$, $P_{VS}$, $Q_{DF}$, $P_{DF}$ are given by \cite{Ohtsuki_2002}. These functions along with the approximated stirring integral can be found in Appendix \ref{sec:stirr}.

A further contribution to the dynamical evolution stems from the stirring {of the planetesimals by the embryos which is given by}
\begin{align}
    \dot{e^2}\big|_{stirr} &= \frac{1}{6} \sum_{j}^{n} f_j \frac{\Omega M_{planete,j}}{6\pi b M_*} P_{VS}(\Tilde{e},\Tilde{i}) \\
    \dot{i^2}\big|_{stirr} &= \frac{1}{6} \sum_{j}^{n} f_j \frac{\Omega M_{planete,j}}{6\pi b M_*} Q_{VS}(\Tilde{e},\Tilde{i}) ,
\end{align}
where the $Q_{VS}$ and $P_{VS}$ are the same functions as above with the caveat that $\Tilde{e} = a_0e_{plan}/R_H$ and $\Tilde{i} = a_0i_{plan}/R_H$ were $R_H$ is the planets hill radius. The distance modulation function $f$ of the $j'th$ planet is given by 
\begin{equation}
    f_j^{-1}= 1 + \frac{|a_0 - a_{planet,j}|}{5R_{H,j}} ,
\end{equation}
and describes the fall of in stirring by the planet outside of its feeding zone. 
For the stirring from the density fluctuation we follow the description of \cite{Ormel_2012} and \cite{Kobayashi_2016} which is given by
\begin{align}
    \dot{e}^2\big|_{DF} &= 400  \alpha  \left ( \frac{H_g a_0 \Sigma_g}{M_*}\right)^2 \Omega + \frac{4\alpha}{3\Omega t_{stop}}\left( \frac{c_s}{\Omega a_0 }\right)^2\\
    \dot{i}^2\big|_{DF} &=  4  \alpha  \left ( \frac{H_g a_0 \Sigma_g}{M_*}\right)^2 \Omega + \frac{2\alpha}{3\Omega t_{stop}}\left( \frac{c_s}{\Omega a_0 }\right)^2 ,
\end{align}
where $H_g$ is the scale height of the gas disk and $t_{stop}$ is the same stopping time as for the damping given in Eq. \ref{eq:tstop}. 

As the gas disk is partially pressure supported, its orbital speed is sub-Keplerian which means the planetesimals experience a headwind when orbiting around the star. Therefore they lose angular momentum which leads to the decay of their semi major axis referred to as drift. This radial motion depends on the same drag regimes \citep{Guilera_2014} as for the damping discussed before and can be described as

\begin{equation}
    \label{eq:v_drift}
    \frac{\partial a}{\partial t} = 
        -\frac{2a\eta}{t_{stop}}  *\frac{s^2}{1+s^2}
\end{equation}
with $s = 2\pi*t_{stop}/P$ and where $P$ is the period and the stopping time is given by

\begin{equation}
\label{eq:tstop}
     t_{stop} = \begin{dcases}
      \frac{6 \rho_p r_p}{\rho_{gas} v_{rel}}  &\text{Quadratic regime} \\
          \frac{2 \rho_p r^2_p}{3\rho_{gas}\lambda c_s} &\text{Stokes regime}\\
          \frac{\rho_p r_p}{\rho_{gas}c_s} &\text{Epstein regime}.
     \end{dcases}
\end{equation}

The consideration of drift is important as we can see the strong dependence of the drift speed on the size of the planetesimals. This becomes important as we reduce the typical size of solids via fragmentation.\\
As the planetesimals drift across the ice line, we expect their volatile components to evaporate reducing the typical mass and also the typical radius of the solids. To account for this we implemented a simple ablation model following the prescription of \cite{Burn_2019}. The ablation follows the theoretic kinetic expression for water ice,
\begin{equation}
    \phi(T) = \frac{P^s(T)}{\sqrt{2\pi m_{H_2O}R_g T}} ,
\end{equation}
where $P^s$ is the water vapour sublimation pressure, $m_{H_2O}$ is the molecular mass of water and $R_g$ is the universal gas constant. The above equation assumes zero partial pressure of water in the vicinity of the planetesimals. From this formula and assuming that the ice is removed from a layer with thickness $\delta \ll r_p$ we calculate the water mass loss, 
\begin{equation}
    \frac{dm}{dt}\Big|_{H_2O} = \phi(T)m_{H_2O}4\pi r_p^2 ,
\end{equation}
with the mass loss calculated we can calculate the change in other properties that result from it like their bulk density {and radius which is done by calculating the density according to the updated ice fraction and computing the new radius from it}. This only affects planetesimals from beyond the initial ice line which have drifted across the dynamic ice line calculated from the structure of the gas disk. This only occurs for small planetesimal sizes ($s<=100m$) because the ice line moves towards the central star faster than the larger solids.

The gravitational interaction between planetesimals not only changes their dynamical state but also leads to radial diffusion. To describe this behaviour we use the prescription of \cite{Tanaka_2003}. The associated viscosity of the diffusion process can be written as
\begin{equation}
    \nu = \sum_{i,j}\frac{1}{12} \langle R_{VSij}\rangle  \mu_{i,j} h_{ij}^4 a_0^4 N_{i}N_j  \Omega/\Sigma_{tot} ,
\end{equation}
where $\langle R_{VS,ij}\rangle = 4/3(\langle Q_{VS,ij}\rangle+\langle P_{VS,ij}\rangle)$ and $N_i$ is the number surface density of the i-th planetesimal swarms, $\mu$ is the reduced mass of the interacting planetesimals and $h_{ij}$ their mutual hill radius. 
We then use the diffusion and the drift velocities to solve the advection diffusion equation for each swarm (including the fragments) which is described by the diffusion advection equation:
\begin{align}
    \frac{\partial}{\partial t}(\Sigma_i) &- \frac{1}{r}\frac{\partial}{\partial r}(r v_{drift}\Sigma_i) - \frac{1}{r}\frac{\partial}{\partial r} \left[ 3r^{0.5}\frac{\partial}{\partial r}(r^{0.5} \nu \Sigma_i)\right] \\
    &= \dot\Sigma_{accretion} + \dot\Sigma_{frag} + \dot\Sigma_{ablation}\nonumber
\end{align}
where on the right hand side, the sink terms describe the mass removal due to the accretion by the planetary embryos, the mass transferred between the swarms due to fragmentation and the ablation of the ices when crossing the water ice line as described above which are treated in separate steps each.
The initialisation of the solid disk is important because it dictates the speed of the early accretion. The initial solid surface density follows a power law with an exponential cutoff described by
\begin{equation}
    \Sigma_0 = \Sigma_{s,0} f_s(r)\left( \frac{r}{5.2\text{AU}}\right)^{-\beta_s} \text{exp}\left(\left(\frac{r}{r_{cut,s}}\right)^{-2}\right) ,
\end{equation}
where the power law is chosen {to be} minimum mass solar nebula-like (MMSN) \citep{Weidenschilling_1977} as $\beta_s = 1.5$ and the outer radius  $r_{cut,s}$ is given as a function of the cutoff radius of the gas $r_{cut,g}$ as $r_{cut,s} = 0.5 r_{cut,g}$ {\citep{Ansdell_2018}}. The initial solid surface density $\Sigma_{0,s}$ is calculated by enforcing a specific dust to gas ratio in the protoplanetary disk. The factor $f_s$ accounts for the fact that not all elements are in the solid phase in all radial parts of the disk. This is calculated with the disk chemistry model of \citep{Thiabaud_2014} and \citep{Marboeuf_2014}. The relative abundances are set according to the interstellar medium (for the volatiles we track H, O, C and S atoms) and thus we calculate the fraction of material that is in its solid phase $f_s$ at each radial location in the disk. This factor becomes unity for large separations past all the ice lines. This also naturally sets the inner solid disk edge when {123} $f_s = 0$ i.e. when the planetesimals sublimate completely. This sublimation radius is enforced, removing the solid mass interior to it throughout the formation stage depending on the temperature structure of the gas disk. The water ice line is calculated in the same way and is the biggest jump in surface density as it makes up $\approx 60 \%$ of the ice mass \citep{Marboeuf_2014}. The bulk density $\rho$ is also determined by the location of the ice line, meaning we consider rocky planetesimals with a bulk density of ($3.2g/cm^3$) inside the initial ice line and icy planetesimals outside with ($\rho =1g/cm^3$).

The initialisation of the dynamical state of the solid disks is simply given by the equilibrium {between} the self stirring of the planetesimals and the gas drag by the gas disk \citep{Fortier_2012}. Note that the initialisation does not change when considering fragmentation as at $t=0$ no fragments are present in the disk and therefore we do not have to consider the dynamic friction term. This results in
\begin{equation}
    e_{plan} = 2.31 \frac{M_{plan}^{4/15}r^{1/5}\rho_{plan}^{2/15}\Sigma_g^{1/5}}{C_D^{1/5}\rho^{1/5}M_*^{2/5}} ,
\end{equation}
and the inclinations are simply given by $\beta = i/e= 0.5$.

\subsection{planet population synthesis}
\label{sec:mc_variables}
In order to probe {the planet formation process for different initial conditions} and to compare our theoretical results with the actual exoplanet population we need to make use of population synthesis \citep{Ida_2004a,Mordasini_2009,Mordasini_2015,NGPPS_2}. The main idea is to run multiple global simulations with different initial conditions to capture the {diversity of resulting planets} and to account for the chaotic nature of the N-body interactions in planet formation \citep{Mordasini_2018}. {For this} we run many systems (typically $\sim1000$) with the \textit{Bern} model as described above where the initial conditions are generated from the following random variables: 
\begin{itemize}
    \item Dust to gas ratio $f_{D/G}$\\
    To constrain the mass of gas in the disk we make use of the gas to solid mass-ratio $f_{D/G}$ and assume the disks and their stars have the same metallicity which leads to
    \begin{equation}
        \frac{f_{D/G}}{f_{D/G,\odot}} =  10^{[Fe/H]} ,
    \end{equation}
    where the metallicities follow the distribution of \cite{Santos_2005}. The entire dust mass is assumed to be converted to planetesimals.
    \item Initial gas disk Mass $M_g$ and dust mass \\
    The distribution of initial dust disk masses $M_g$ reproduces the Class I disks reported in \cite{Tychoniec_2018}. The dust masses of the disks follow a log normal distribution with $\mbox{log}_{10}(\mu/M_\oplus) = 2.03$ and $\sigma = 0.35\mbox{dex}$. The gas mass is then obtained by dividing by the previously defined dust to gas ratio. The sampled gas disk masses range is limited from $0.004 M_\odot$ to $0.16 M_\odot$ to ensure self-gravitational stability. 
    \item Photo evaporation rate $\dot{M}_{wind}$\\
    Since the disk lifetimes are constrained by a combination of $\alpha$ and $\dot{M}_{wind}$ and $\alpha$ is {kept constant at a value of $2\times 10^{-3}$}, we chose the evaporation such that the disk lifetimes fit observations \citep{NGPPS_2}. This results in a log-normal distribution with $\text{log}(\mu/(M_\odot/yr))=-6$ and $\sigma = 0.5$dex.
    \item Inner edge of the gas disk $r_{in}$\\
    The inner disk edge {of the gas disk} is chosen to be the co-rotation distance to the star i.e. where the Kepler period matches the rotation period of the star. The stellar periods are sampled from \cite{Venuti_2017}.
    \item Initial location of the embryos $a_{embryos}$\\
    The initial embryo locations are randomly chosen between the inner disk edge and $40$ Au {uniformly in logarithm}. To reproduce the results from N-body simulations \cite{Kokubo_2002}, the embryos can not be closer to another than $10$ hill radii if we consider multiple embryos.
\end{itemize}
In addition the remaining parameters can be found in table \ref{tab:param_syn} in the appendix. Note that all of the systems are formed around solar mass stars. One particular parameter of interest is the chosen planetesimal size which has a lot of effect on the fragmentation model introduced in the next section. We consider two choices for the planetesimals size $100km$ and $1km$. The first choice is motivated by the size distribution of asteroids in the main belt \cite{Morbidelli_2009} and the results from streaming instability \cite{Schaefer_2017}. The second choice lets us explore setting closer to the previous choice of \cite{NGPPS1} which is motivated by \citep{Schlichting_2013,Arimatsu_2019,Kenyon_2012,Weidenschilling_2011}.
 
\subsection{Fragmentation model}
\label{sec:toy}
To study the influence of fragmentation on the population of exoplanets we have {to use a simplified} model that follows in parts the one described in \cite{Ormel_2012} with a few notable differences. The model described in \cite{Ormel_2012} is a local model and does not support multi planet systems and migration. Furthermore it takes a simplified approach to the calculation of the dynamical state and the effects of drift. {Our} model adds to the description of the solid disk a swarm of solids of a second, variable size named fragments that are created by the mutual collisions of planetesimals.

First we will take a look at {the baseline description of the} the fragmentation model. In this case the fragments are modelled as an additional swarm of solids of a fixed size ($s_f$) which is a parameter of the model. At each  time step we calculate the fragmentation rates from the mutual collisions of the planetesimals (PP) and if there are any, the eroding collisions of planetesimals with fragments (PF). At each radial bin we calculate first the collisional time scale among planetesimals which is given by

\begin{align}
    \label{eq:t_coll}
    T_{coll\: i,k} = \frac{1}{\sigma_{cs}*n_{k}*V_{rel}} ,
\end{align}

where $V_{rel}$ is the relative velocity among the colliders. The number density $n_{k}$ of projectiles and the cross-section are given by,
\begin{equation}
    n_k = \frac{\Sigma_k/\frac{4\pi}{3}\rho_k s_k^3}{2h}\:\: , \sigma_{cs} =\pi(s_i+s_k)^2*f_f ,
\end{equation}
where $\rho$ is the bulk density of the planetesimals, $h = a*i_{max(i,k)}$ is the scale height of the solid disk and $f_f$ is the gravitational focus factor \citep{Morbidelli_2009}. The subscripts $i$ and $k$ refer to the properties of the targets and projectiles respectively. When we consider collisions among different swarms the scale height used is the maximal one as it reflects the true number densities in the collisional volume. {The gravitational focus factor describes the deviation of the collisonal cross section from the geometric one i.e. it accounts for the fact that the colliders have a gravitational field enhancing their cross section.} For this we follow the description of \cite{Morbidelli_2009}. We define the ratio of specific collision energy and the specific material strength as follows, 
\begin{align}
    q_{i,j} &= \frac{0.5m_\mu v_{rel}^2}{(m_i+m_j)Q_{d,i}^*} \\
    q_{i(,i)} &= \frac{\Delta v_{ii}^2}{8Q_{d,i}^*} ,
\end{align}
where $m_\mu = (1/m_i + 1/m_j)^{-1}$ is the reduced mass. For the specific fragmentation energy $Q_d^*$ we use the one from the bigger collider ($i$) which is considered to be the target. The excavated mass of a collision $m_{exc}$ is then given by the collisional outcome model of \cite{Kobayashi_2010}, 
\begin{align}
\label{eq:m_ex}
    m_{exc} = \frac{q_{i,j}}{1+q_{i,j}} (m_i+m_j) .
\end{align}

We assume that all of the mass that gets excavated flows instantaneously to the fragments. This is {justified by the fact that} the bodies in the gravity regime have lower material strength with reduced size, which should induce a collisional cascade and the size of the fragments represents its lower end \citep{Ormel_2012} {(see or discussion in section \ref{sec:limitations})}. In the nominal case we do not consider {a change in the typical size of planetesimals} due to these collisions for the planetesimals {despite the mass that is excavated}. {This is ignored because outside of the feeding zone of embryos the collisional evolution is in the runway regime due to the lack of added stirring by larger bodies. This, combined with the fact that we do not track individual bodies and only consider one size of planetesimals means we are not able to resolve the behaviour of the runaway bodies.} This could then affect the growing embryo as it migrates. The resulting change in surface density from the collisions is then obtained by combining Equations (\ref{eq:t_coll}) and (\ref{eq:m_ex}) which results in 
\begin{equation}
        \label{eq:sigmadot}
        \dot{\Sigma}_p = -\frac{q_p}{1+q_p} \frac{\Sigma_P}{\text{T}_{collPP}} - \frac{q_{pf} \Sigma_P}{\text{T}_{collPF}} = -\dot{\Sigma}_f ,
\end{equation}
 
where the second term stems from the collisions with fragments. Note that because the impact energy of the fragments is significantly smaller $q_{PF} \ll 1$ we can use that $q_{PF}/(1 + q_{PF} ) \approx q_{PF}$.

\begin{figure}
    \centering
    \includegraphics[width=\hsize]{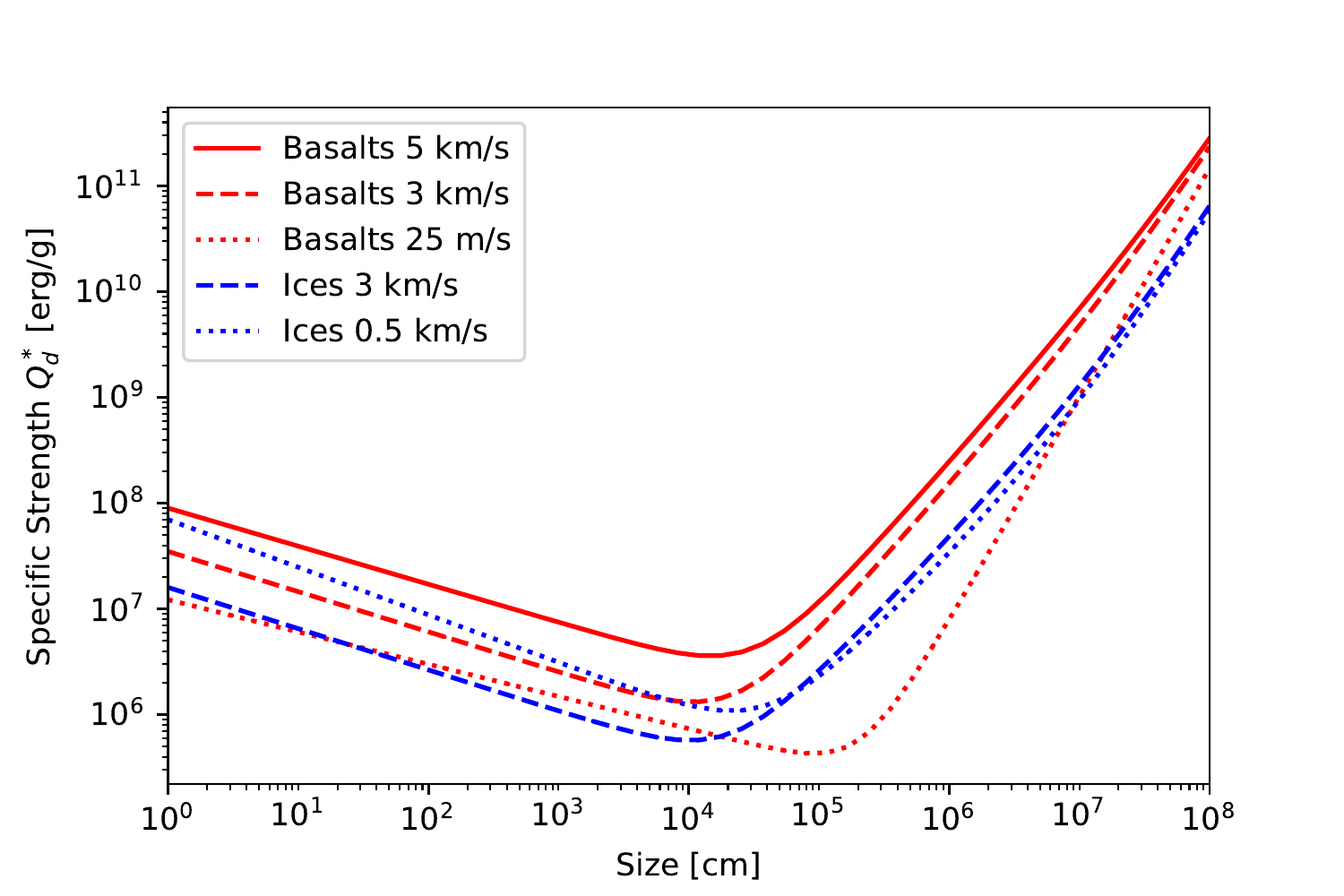}
    \caption{The specific material strength for different ice (blue) and basalt (red) at different sizes and impact velocities}
    \label{fig:Qd_*}
\end{figure}

The description of the material strength is vital for the fragmentation model because it dictates the speed of fragmentation and the size distribution of fragments. We utilize the specific fragmentation energy $Q_d^*$ i.e. the energy needed to disperse 50\% of the target body as it is used to determine the collisional outcomes as described above. For this model we consider two descriptions of $Q_d^*$. The nominal {case takes the results} from \cite{Benz_1999} at $v = 3km/s$ for basalt and ice respectively as described below, 
 \begin{align}
    Q_d^*(s) = Q_{0s} \Big(\frac{s}{cm}\Big)^{b_s} + Q_{0g}\rho_s\Big(\frac{s}{cm}\Big)^{b_g} + 9v_{esc}^2(s) .
\end{align}
Where $s$ refers to the size of the planetesimals and $v_{esc}$ is their mutual escape velocity {and the remaining parameters can be found in \ref{tab:Qd}}. The first term refers to the strength regime, the second term is the gravitational regime and the last term is the gravitational potential \citep{Stewart_2009}. A visualisation of the function can be seen in Fig. \ref{fig:Qd_*}. {We also consider another description} of the fragmentation energy $Q_d^*$, which takes into account the effect of different impact velocities and the differences in strengths between icy and rocky planetesimals. Specifically this means that we interpolate between the curves given by the parameters in the table \ref{tab:Qd} in the appendix according to the impact speed and ice fraction of the planetesimals. The parametrisations of $Q_d^*$ are described in \citep{Benz_1999} and \citep{Benz_2000} and the interpolation follows the approach of \cite{San_Sebastian_2019}. {For velocities} above or below the maximum/minimum relative velocity {covered by the parameters in table \ref{tab:Qd}}, we take the values of the closest curve, i.e. we do not extrapolate beyond the curves. This leads to a more accurate representation of the specific material strength which mainly impacts the low velocity regime where the material strength is considerably lower.

The fragment size can be either given as a parameter or can be calculated. When calculated by the model, the initial fragment size is given by the minimum in the specific fragmentation energy $Q_d^*$ for the respective materials. This means icy and rocky planetesimals have different initial sizes. This leads (for an ice fraction of: $\approx0.6$) to an initial fragment size of $105\:m$ inside the ice line and a size of $147\: m$ outside. The size of fragments is then only changed if the collisions between the fragments become destructive i.e. if $\Delta v_{FF}^2/8Q_d^*(s_f)>0.5$ meaning the collisions between the equal sized fragments excavate more mass then they add. In this case We reduce the size of fragments until the collision become non destructive, meaning their size at each radial separation of the disk is then given by the implicit equation
\begin{align}
\label{eq:f_size}
    \frac{\Delta v_{FF}^2}{8Q_d^*(s_f)} = 0.5 .
\end{align}
In this equation $s_f$ is the fragment size and $\Delta v_{FF}$ is the relative velocity among fragments. Due to the negative slope in the strength regime of the $Q_d^*$ curve and the increased gas damping at lower sizes we are able to find a stable size for the fragments i.e. Eq. \ref{eq:f_size} has a solution. 

{The fragments have short drift timescales due to their strong coupling to the gas. As dictated by the pressure structure of the gas disk they will drift towards the central star until they reach a pressure maximum. This leads to a significant pileup of fragments at the global pressure maximum located at the inner disk edge \citep{Guilera_2017}. For our gas disk structures and condensation model \citep{Thiabaud_2014} this pressure maximum is located outside of the dust sublimation line. The resulting pileup of fragments can lead to very high solid to gas ratios at the pileup which is not well described by our model as we do not account for the back reaction of solids onto the gas. Furthermore we expect mass to be removed from the inner disk edge through various processes \cite{Li_2021} however the exact structure of the inner disk edge remains quite uncertain \citep{Dullemond_2010}. As we are not able to resolve the full dust evolution and collisional evolution that comes with the pileup we use a heuristic simplified treatment of the solids at the global gas pressure maximum. We limit the solid surface density to the gas surface density when the gas surface density is above a certain threshold (which we chose to be $200g/cm^2$). We ignore the limiting at lower gas surface densities because we do not expect solids in the entire disk to vanish when the gas disk dissipates and the drifting of solid slows down significantly as the gas disk dissipates. We discuss this simplified treatment and its effects in Sect. \ref{res}.}

The code features an adaptive time step for the solid disk that ensures that the processes are treated consistently which follows the approach of \cite{Morbidelli_2009}. The time step is chosen such that neither the surface density ($\Sigma$) nor the random velocities ($e$ and $i$) change by more then $10\%$ anywhere in the disk in a single time step. {Another} limit is that we ensure that the solids especially the fragments do not drift farther then $1\%$ of their semi-major axis but there is a minimal time step of $1$yr. Note that these restrictions for the timestep are {additional} to the ones described in \cite{NGPPS1} {that come from the planets internal structure and growth and the evolution of the gas disk.}

\section{Comparison with previous work}
\label{test}

In order to ensure the validity of the fragmentation model introduced in the last section it is important to make a comparison to similar models presented in the literature namely \cite{Chambers_2006} and \cite{Ormel_2012}. Those models both operate with a similar 3 component approximation to the solids in the disk (embryos, planetesimals and fragments).

First we want to compare the fragmentation rates of our planetesimal planetesimal collisions to the ones from \cite{Chambers_2006} which are given by

\begin{equation}
    \frac{\text{d}\Sigma_p}{\text{d}t} = \frac{69.4 \Sigma_p^2 a_0^2}{Pm}*\text{min} \left[\frac{Q}{2Q_d^*},1\right] .
    \label{eq:chambers_frag}
\end{equation}

\begin{figure}
    \centering
    \includegraphics[width=\hsize]{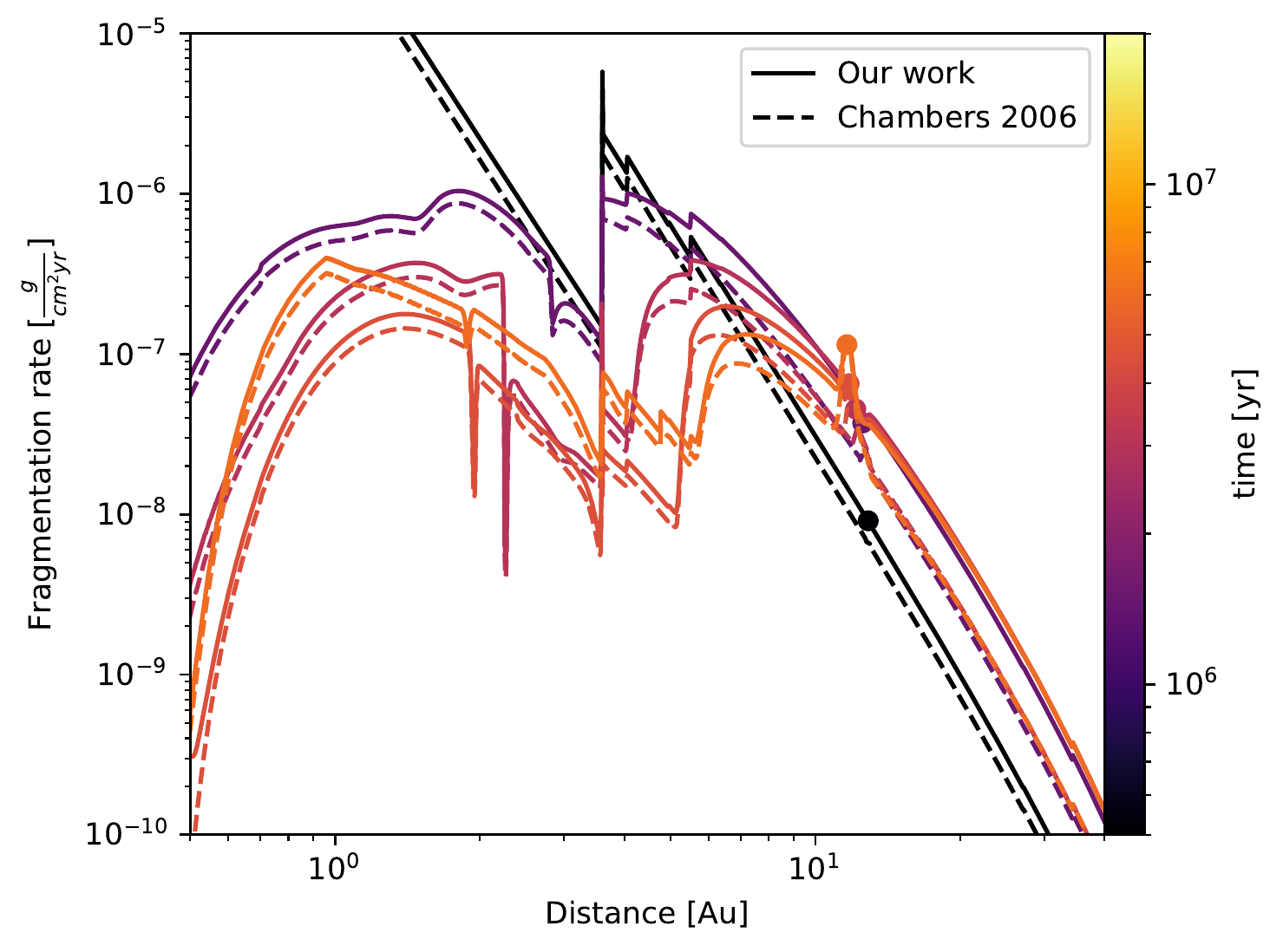}
    \caption{Fragmentation rates of the planetesimals due to their mutual collisions in our model (solid) and from Eq. \ref{eq:chambers_frag} (dashed) at different times. The dot marks the forming planets location}
    \label{fig:chambers_comp}
\end{figure}

Where Q is the specific impact energy i.e. $Q = 0.5*m_\mu*v_{rel}/m_p$. This formula is constructed under the assumption that $\beta = i/e = 0.5$ which does not hold throughout the entire disk in our simulations. The comparison of our fragmentation rates {resulting from planetesimal planetesimal collisions} and theirs can be seen in Fig \ref{fig:chambers_comp}. {We test this on a system with }initial condition that can be seen in table \ref{tab:NGPPS11_init} in the appendix.

As we can see the rates are quite close to one another except on the outer edges of the feeding zone of the embryo which is expected as this is where we get values of $\beta$ that are significantly lower then the equilibrium value of $\beta = 0.5$. This arises due to the stirring of the planet that affect the eccentricities and inclinations with different strength leading to a deviation from the equilibrium value of $\beta = 0.5$. Furthermore \citep{Chambers_2006} assumes all the collisions to happen in the high velocity regime which might not hold far away from the embryo.

For the second comparison we construct a test system with the initial conditions that can be seen in Table \ref{tab:param_test}. {Additionally, To be consistent with the works of \cite{Ormel_2012}, for the profile of the disk we chose MMSN-like initial conditions of}

\begin{align}
    \Sigma_{gas}(a) = 376.2 g/cm^2  \left(\frac{a}{5.2\text{AU}}\right)^{-1.5}\\
    \Sigma_{planetesimals} = 1/57 \Sigma_g(5\text{AU}) = 7 g/cm^2 .
\end{align}

This corresponds to a MMSN model with an enhanced surface density of $2.65$. Furthermore we set the fragment size to a fixed size of $1cm$ and planetesimal are set to a size of $10^6cm$ so that the model can be compared to Fig. 6 from \cite{Ormel_2012}. {Additionally we disable planet migration, the evolution of the gas disk and the drift of solids and the gravitational focusing for the collisions as they do not consider these processes. For the description of the specific fragmentation energy we use $Q_d^*$ description of \cite{Benz_1999} for ices at $500 m/s$ (see table \ref{tab:Qd}). For the relative velocity between solids for this comparison we also follow their simplified formula of $\Delta v_{PP} = \Delta v_{PF} = 2e_P *v_k$. To have a comparable treatment for the dynamical state we increase the integration time of Eq. \eqref{eq:dynamical_evolution} by a factor of 50 which ensures the dynamical state is always in equilibrium.}

\begin{figure}
    \centering
    \includegraphics[width=\hsize]{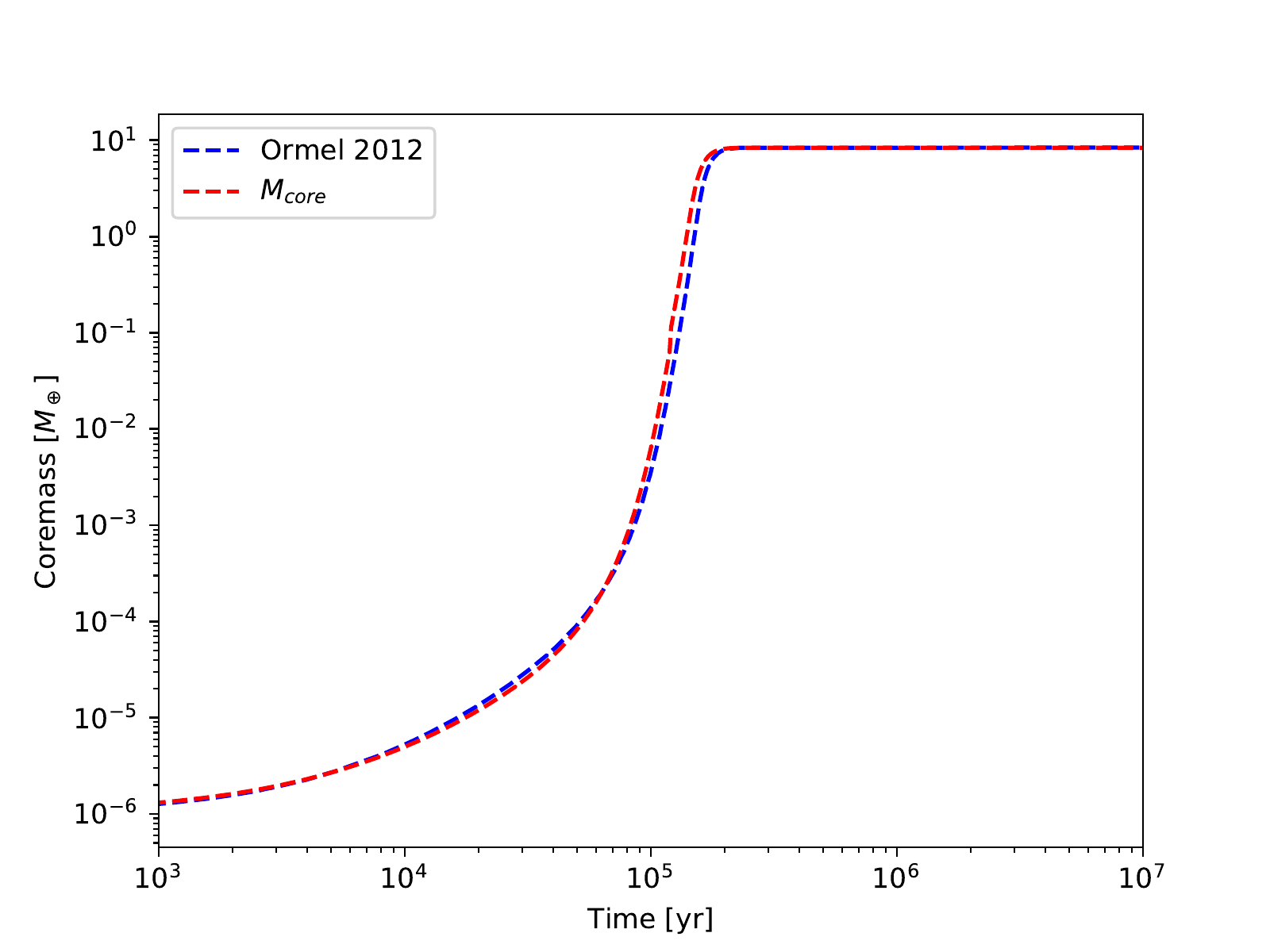}
    \caption{Core mass of the embryos in our simulation (red) and the reference one from \cite{Ormel_2012} Fig. 6A (blue)}
    \label{fig:ormel_growth}
\end{figure}

The resulting mass growth of the embryo is displayed in Fig. \ref{fig:ormel_growth}. The major difference between the two runs is in the initial stages of embryo growth which is mainly due to the different ways the eccentricities and inclination are calculated, they consider the balance between the two dominant timescales whereas we integrate the dynamical evolution equation given in Eq. \ref{eq:dynamical_evolution} which leads to different eccentricities and inclinations during the early stages. However the end result is very close to one another as the isolation mass is reached almost simultaneously.

\begin{figure}
    \centering
    \includegraphics[width=\hsize]{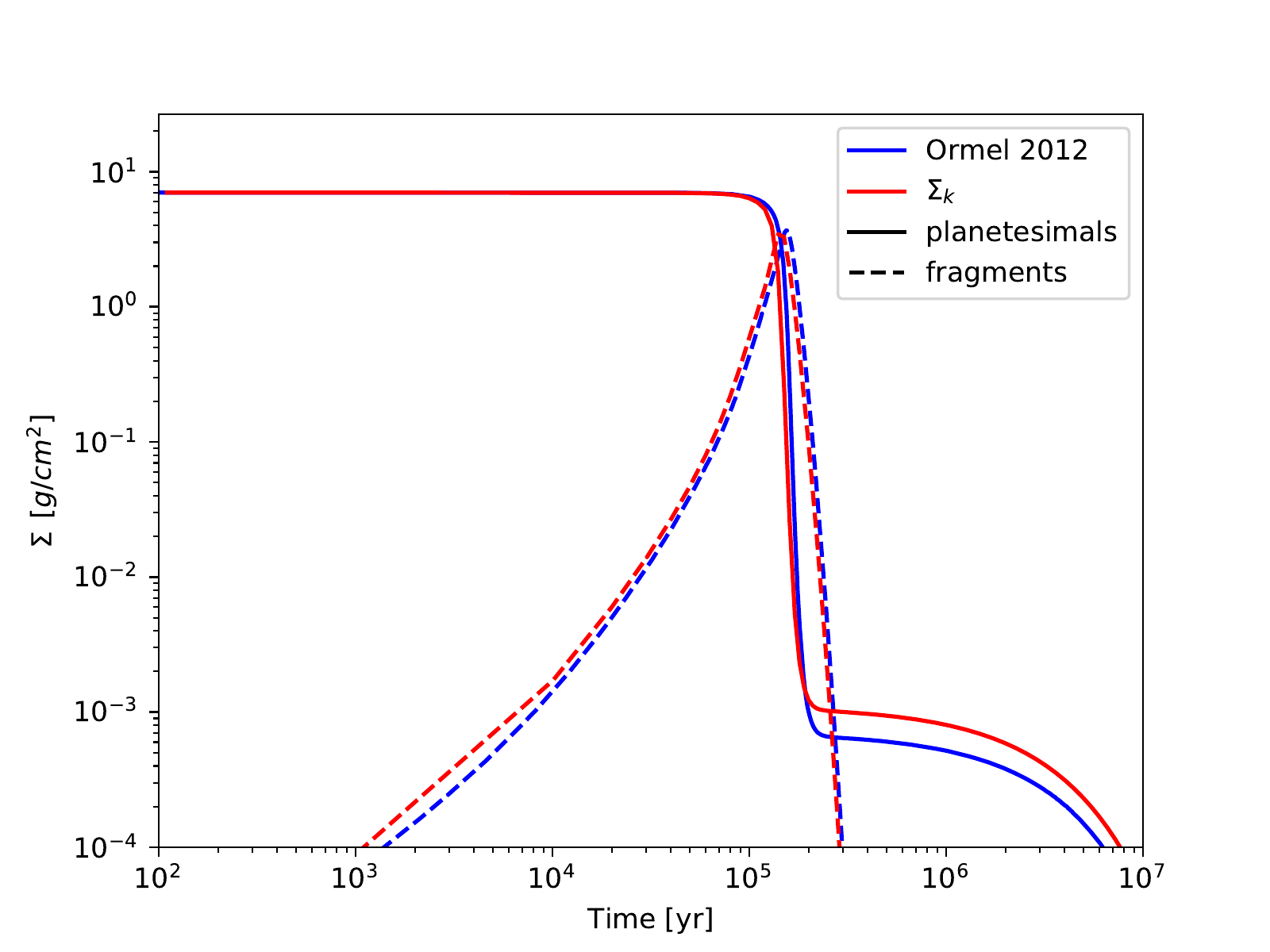}
    \caption{Surface densities of the Planetesimals (solid) and fragments (dashed) at the embryos location for our model (red) and the model of \cite{Ormel_2012} (blue)}
    \label{fig:ormel_surface}
\end{figure}

In addition to the {evolution of the forming planet} we can {compare} the surface densities of planetesimals and fragments at the embryos location which ultimately dictates the accretion rates. The surfaces densities at the embryos throughout the simulation can be seen in Fig. \ref{fig:ormel_surface}

The surface densities around the embryos compare rather well, with the biggest differences {being present} in the early stages of the formation process which can once again be explained by the different way the dynamical state of the solids is calculated, which mostly comes into play at early times. As we can see the {outcome of our} model is comparable to the results of different simplified models used to characterise fragmentation.

\section{Population synthesis}
\label{pop}
To get a more robust understanding of how our newly added fragmentation model affects the synthetic populations we consider {different} setups for the treatment of the solid disk. To do this we ran {single embryo} populations for three different models for two different sizes of planetesimals: $100km$ and $1km$. The three different setups differ in their treatment of fragmentation. The first one represents a reference case that does include the updated stirring, drift and treatment of the ice line for the solid disk but {not} fragmentation. The second one includes fragmentation with a fixed fragment size of $100m$. Third one adds the dynamical calculation of the fragment size as dictated by Eq. \ref{eq:f_size}. {to investigate the formation of multiple planets in the same disk} we completed a population with 20 embryos for {both planetesimal sizes of $100km$ and $1km$ that do consider fragments of $100m$. We also considered the dynamical treatment of the fragment size for the $100km$ planetesimals. As a comparison we also ran} non-fragmenting populations with 20 embryos for the $100km$ and the $1km$ planetesimals. {An overview of} the populations with their abbreviations and their chosen model parameters can be seen in Table \ref{tab:Pops}.

\begin{table}
    \caption{Chosen settings for the different populations}
    \centering
    \begin{tabular}{c c c}
        \hline
        \hline
        Name & planetesimal size & fragment size  \\
        \hline
        $L_{no}$ & $100km$ & no fragmentation \\
        $L$ & $100km$ &  $100m$  \\
        $L_{dyn}$ & $100km$ & dynamic  \\
        \hline
        $S_{no}$ & $1km$ & no fragmentation \\
        $S$ & $1km$ & $100m$ \\
        $S_{dyn}$ & $1km$ & dynamic \\
        \hline
        $ML_{no}$ (20 embryos) & $100km$ &  no fragmentation \\
        $ML$ (20 embryos) & $100km$ &  $100m$  \\
        $ML_{dyn}$ (20 embryos) & $100km$ &  dynamic  \\
        \hline
        $MS_{no}$ (20 embryos) & $1km$ &  no fragmentation  \\
        $MS$ (20 embryos) & $1km$ &  $100m$  \\
    \end{tabular}

    \label{tab:Pops}
\end{table}

\subsection{single embryo populations}\label{sec:single}
    
In order to isolate the effects of fragmentation it is useful to discuss the formation of single embryos in the disk as it allows us to prevent any chaotic noise stemming from the N-body interaction between the different forming planets. For this we investigated different settings for the fragmentation model $L$ and $L_{dyn}$: $L$ is the nominal case with $100km$ planetesimals and a constant fragment size of $100m$. The fragment size was chosen to be close to the minimum in the $Q_d^*$ functions for ice and basalt from \cite{Benz_1999}. In the second population $L_{dyn}$ we also enable the dynamical calculation of the fragment size as described in Sect. \ref{sec:toy}. We also ran a population without fragmentation to have a reference case to compare it to {($L_{no}$)}. As the effects of fragmentation strongly depend on the size of the initial planetesimals we {also run simulations} with smaller planetesimals of $1km$ size in the set of populations $S_{no}$, $S$ and $S_{dyn}$ where we explore the same three settings of our fragmentation model.

\begin{figure*}
    \centering
    \includegraphics[width=0.8\hsize]{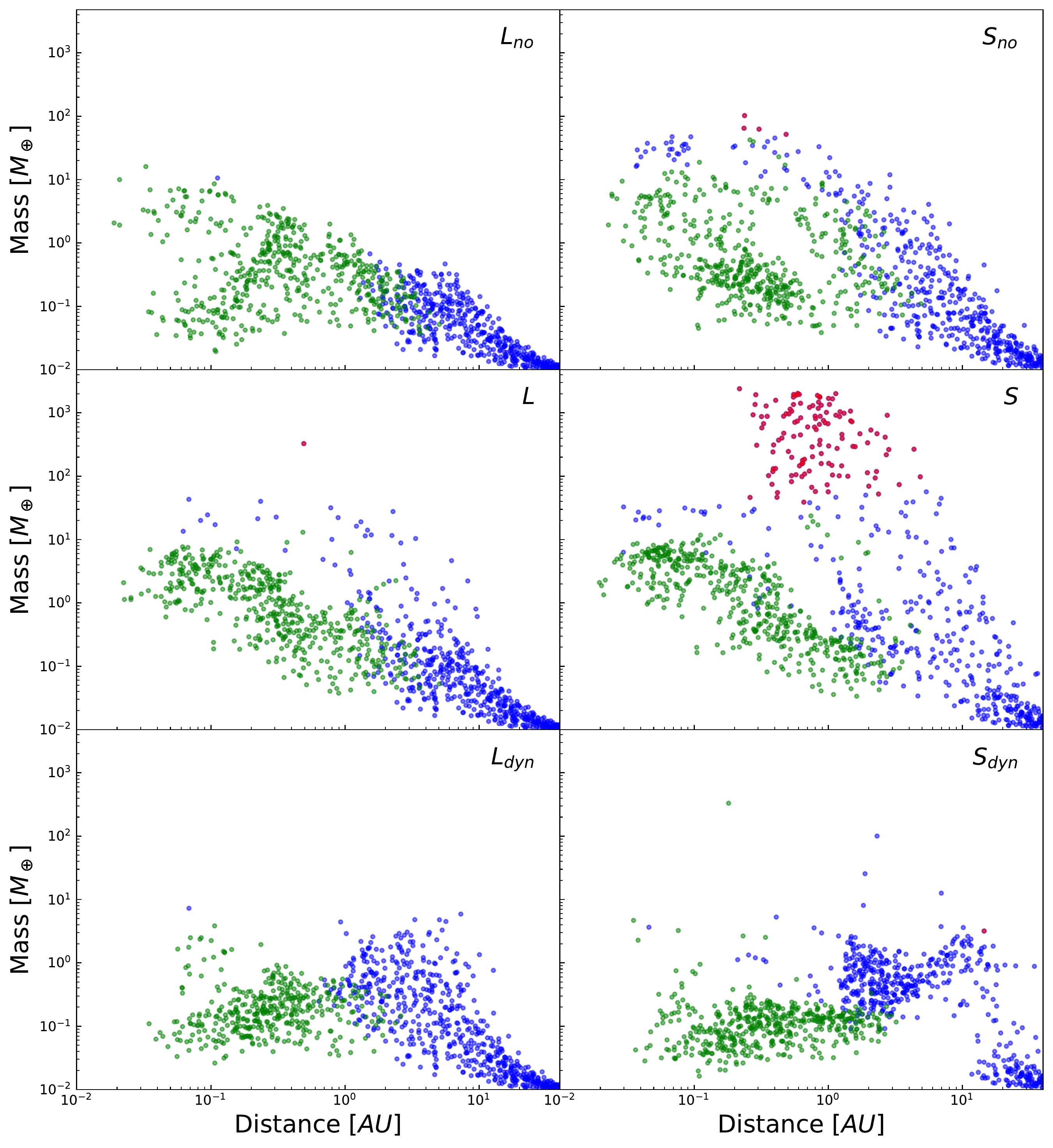}
    \setlength{\belowcaptionskip}{-8pt}
    \caption{The distance Mass diagrams for the populations with a single embryo. The colour of the points refer to their bulk composition: red ones have $M_{env}/M_{core}>1$ while the blue points have a volatile fraction of $>1\%$ and the remaining points i.e. rocky planets are green.}
    \label{fig:sma_m}
\end{figure*}

One of the Main results of the population synthesis is the semi-major axis mass diagram as it allows us to see where the forming planets end up and their formation pathways to their final location. To illustrate this we show the semi-major axis mass diagram for our different single embryo populations in Fig. \ref{fig:sma_m} where the blue markers refer to planets that have accreted more than 1\% of volatile material called icy planets and red ones have a higher envelope then core mass i.e. gas giants. The remaining planets are considered to be rocky and have green markers. All of the following diagrams show their respective populations at a time of $5Gyr$.

\begin{figure}
    \centering
    \includegraphics[width=\hsize]{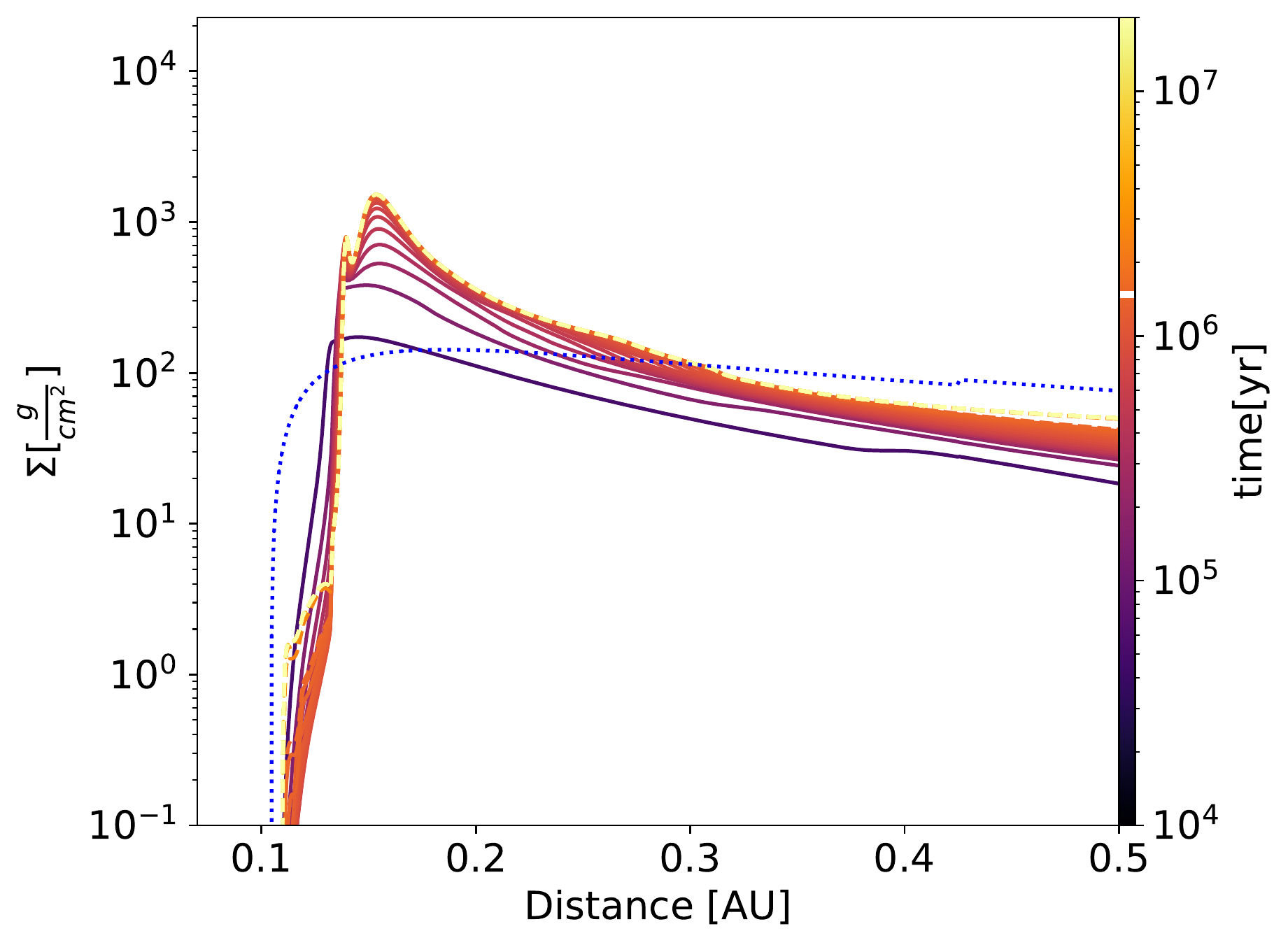}
    \caption{The surface density evolution of the fragments of an example disk with the initial solid surface density in blue and the disk lifetime is marked in white. The dashed lines mark the evolution after disk dispersal.}
    \label{fig:inner_pileup}
\end{figure}

In the left column of Fig. \ref{fig:sma_m} (i.e. for the $100km$ sized initial planetesimals) when comparing the population $L$ with the reference case $L_{no}$ we can see a few notable differences. Firstly we can observe generally more massive planets when we do consider the effects of fragmentation. This is especially visible for the icy planets where for the fragmenting population we see the vertical branch of inward drifting icy planets at around $\sim 10M_\oplus$ as described by \cite{Mordasini_2012,Mordasini_2009}. Furthermore we get significantly less very low mass planets of around $\sim 0.1M_\oplus$ in the inner regions of the disk. This feature mainly arises due to the transport of fragments to the inner disk where they pile up and enhance the growth rates. To illustrate this pileup we show an example system where we omitted the limiting of the surface density described in section \ref{sec:toy} as this makes it easier to see the mass transported to the inner disk. This can be seen in Fig. \ref{fig:inner_pileup} where we plot the surface density evolution of the fragments close to the star. As we can clearly see there is a significant enhancement of the solid surface density of fragments at the inner disk edge when compared with the initial solid surface density. The initial disk profile represents the distribution of solids in the non fragmenting case as the drift of $100 km$ planetesimals is negligible. We additionally see a void of earth mass planets at $\sim 1AU$ in population $L$ which will be discussed later.

When we include the dynamic calculation of the fragment size in population $L_{dyn}$ we similarly see that the growth in the outer disk is greatly enhanced. This holds true even further out in the disk when compared to population $L$. Additionally, we observe icy planets being formed closer to the star down to $\sim 1AU$. This is true even for low mass planets ($M<M_\oplus$) that experience little migration. This can be explained by the drift timescales of the smaller fragments that are significantly shorter leading to increased volatile transport {to the inside of the} initial ice line. The population in the inner disk is significantly less massive due to the short drift timescale of the solid material in the inner disk. {Indeed,} this material is not entirely available to be accreted due to the limiting of the surface density at the inner disk edge as described in Sect. \ref{sec:toy}. 

When we look at the $1km$ sized planetesimals in the right column of Fig \ref{fig:sma_m}, (populations $S_{no}$, $S$ and $S_{dyn}$) the effects remain largely the same as before but with the caveat that the planetary masses are already enhanced due to the faster accretion of the smaller planetesimals. This allows giant planets to be formed in the population with fragmentation whereas they are absent in the reference simulation $S_{no}$. Additionally we once again see the lack of planets forming at earth's location for the fragmenting population. When including the dynamical fragment size we once again see the same imprint as for the bigger planetesimals mainly the halted growth for rocky planets and enhanced growth outside the ice line. Interestingly we see a sharp transition in planet masses at $\sim 10$ AU separating embryos experiencing no growth with $M \approx 10^{-2}M_\oplus$ an the more massive ones.

When comparing the populations $S_{no}$ to the previous populations computed with the previous iterations of the code shown in \cite{NGPPS_2} we observe generally a significantly less massive population. Specifically, in the $S_{no}$ populations no giant planets are formed. However there are three main differences in the updated model that inhibit the formation of giant planets. {Those are the choice of planetesimal size which is increased from $300m$ to $1km$ and the inclusion of stirring of the planetesimals by density fluctuation which both lead to higher eccentricities and inclinations for the planetesimals making them harder to accrete. Additionally, we consider different distributions of the initial conditions, which results the shorter lifetimes of the gas disks (see Sect. \ref{sec:mc_variables}) leaving less time for gas giants to be formed.} 

\begin{figure*}
    \centering
    \includegraphics[width=0.8\hsize]{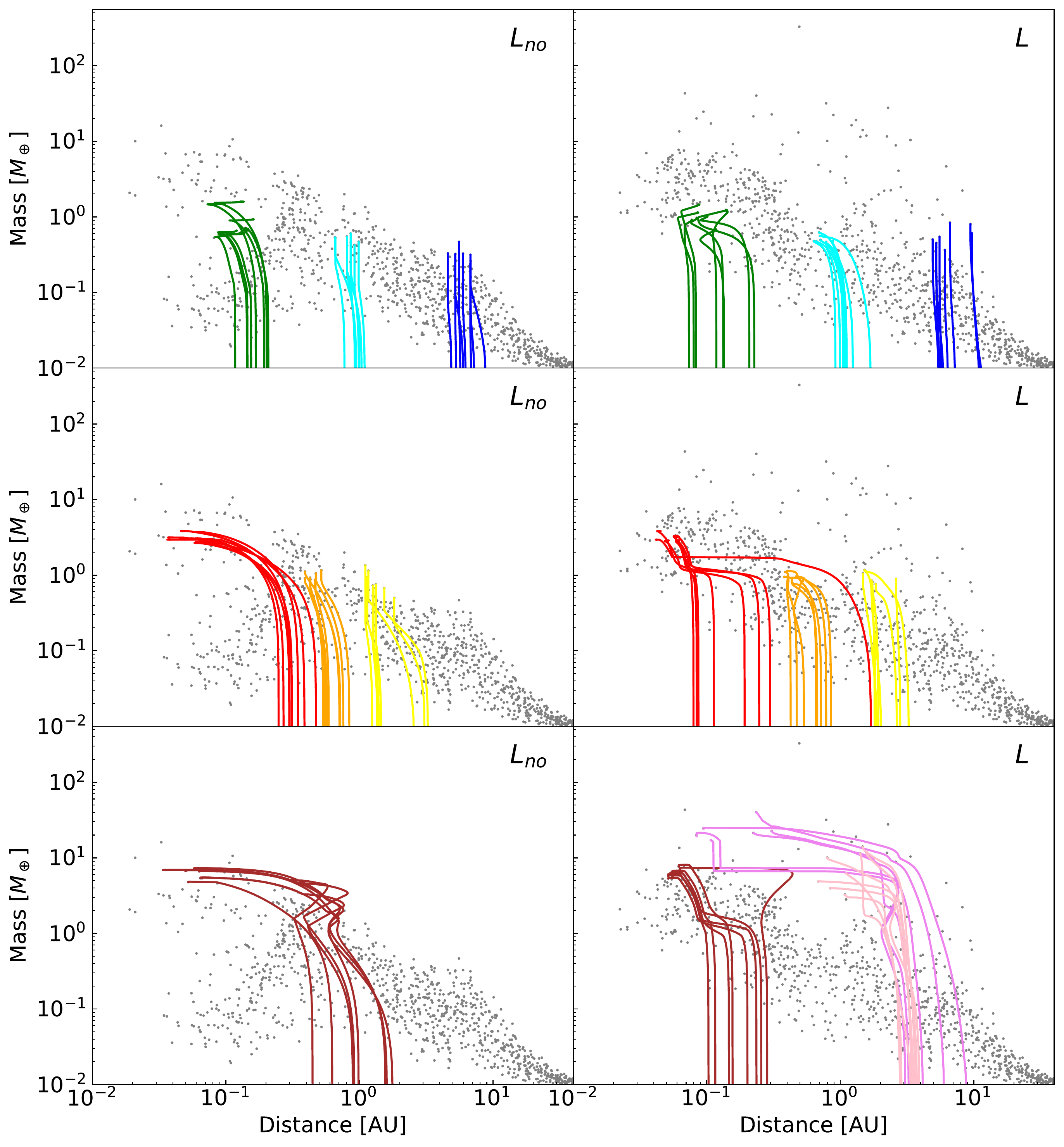}
    \setlength{\belowcaptionskip}{-8pt}
    \caption{The formation tracks (of populations $L_{no}$ (left) and $L$(right)) of 9 groups of systems that form similar planets with different colors. The groups are defined by the final distance and mass of the formed planets and are explained in the text. Each row probes the formation planet types with increasing final mass at different separations from the star}
    \label{fig:track_groups}
\end{figure*}

A planets' properties not only depend on their final location but also their formation pathways. Therefore we are also interested in the formation tracks for different types of planets forming in the populations. Tracks for different types of planets forming in the populations $L_{no}$ and $L$ can be seen in Fig. \ref{fig:track_groups}. This is done by plotting the formation pathways of a group of planets that have similar final properties i.e. they end up at the same location in the distance mass diagram. To determine what is close we use the logarithmic distance in the semi major axis mass plane $d(i,j) = \sqrt{log(m_i/m_j)+log(a_i/a_j)}$ where $m_{[i,j]}$ are the final planet masses and $a_{[i,i]}$ the semi major axes. The final planet properties in [AU,$M_\oplus$] for the groups we chose for Fig. \ref{fig:track_groups} along with their chosen color (that have no physical meaning) are: green:(0.1,1), cyan(1,0.5), blue:(10,0.5), red:(0.05,3), orange:(0.5,1), yellow:(2,1), brown:(0.05,7), pink:(1,7),violet:(0.2,30).

The groups that are not shown {in either of the columns} of  Fig. \ref{fig:track_groups} have no planet that is in close enough proximity ($d_{min}>2$) to the groups centre i.e. these particular types of planets do not form in that population {(for example the violet group in the bottom left panel)}. The formation tracks for the first row of Fig. \ref{fig:track_groups} look quite similar between the two populations with exception for the green group where the fragmenting population keeps accreting material during outwards migration. {The second row however displays interesting differences when it comes to the red group where in the fragmenting population the planets accrete a larger part of their mass closer to the star when compared to their non fragmenting counterparts}. The planets growing further out (yellow and orange) form almost in situ meaning they migrate very little during their formation. But when comparing the orange and cyan groups we see that for population $L$, the planets show more migration close to their final mass where as for population $L_{no}$ the planets migrate during the entire growth process, which implies difference in timing of the accretion. The void of planets at earths location seen in the population $L$ can be explained by the difference in migration where we see the yellow group experiencing little migration and the orange group migrating significantly more leaving this part of the diagram depleted. The migration becomes more significant when we get to $\sim 10$ earth masses which is an expected result for type I migration \citep{Ward_1997}. When comparing the brown group we see a similar picture as for the red one, The fragmenting population accretes more of its material closer to the star where the same type of planet migrating inwards forms further out in the disk in the non fragmenting case. The purple and pink groups are absent from population $L_{no}$ as the planets beyond the ice line don't grow massive enough to fill that part of the diagram {i.e. there are no planets forming with $M > M_\oplus$ beyond 1 $AU$ for population $L_{no}$}. 

\begin{figure*}
    \centering
    \includegraphics[width=\hsize]{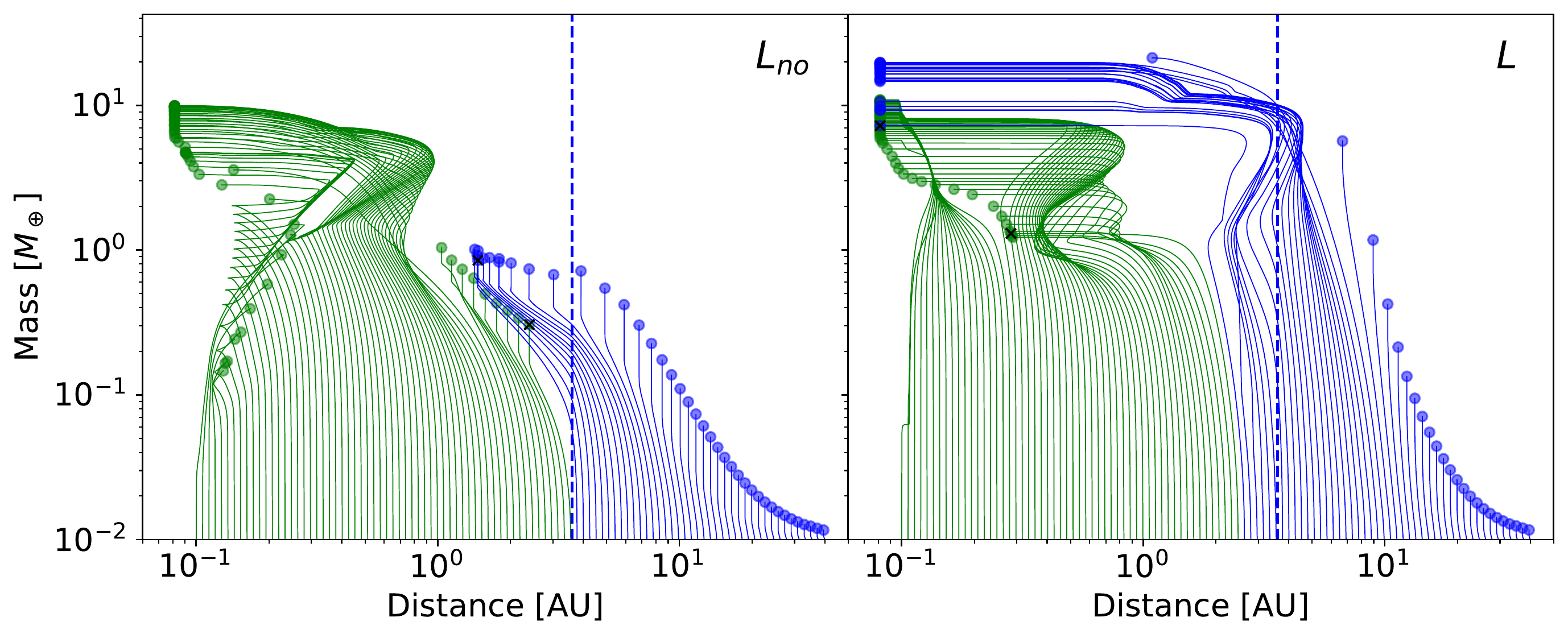}
    \caption{The tracks for 100 separately evolved embryos at different starting locations with the same initial conditions. On the \textit{right} the model choice is the same as population $L$ and on the \textit{left} the model choice is the same as population $L_{no}$. The colour separates icy (\textit{blue} $>1\%$volatiles) from rocky planets (\textit{green}) and the initial water ice line is plotted in blue (dashed). {The furthest initial embryo forming a rocky planet and the closest initial embryo forming an icy planet have bee marked with a black x}}
    \label{fig:trajectories}
\end{figure*}

We can also probe the importance of the starting location of the embryo by running different simulations of the same initial disk (see parameters in \ref{tab:NGPPS11_init}). To probe the importance of the initial location of the embryo we run 100 single embryo simulations with different starting locations. We do this with and without the fragmentation of planetesimals to be able to compare the two scenarios. This can be seen in Fig \ref{fig:trajectories}, note that the embryos are equally spaced in log between the inner edge of the disk and 40 AU as this is also the range of semi major axes allowed for the seeding of the initial embryo location for our populations and it has the same spacial density distribution {(uniform in log)}.
    
When we look at the non fragmenting tracks of Fig. \ref{fig:trajectories} we can see that all of the icy planets originate from beyond the initial ice line because the 100km planetesimals experience very little drift. For the rocky planets we see that most of them end up at the inner disk edge and that the more massive ones originate from further out in the disk which is consistent with the brown group from Fig. \ref{fig:track_groups}. The tracks in the right panel of Fig. \ref{fig:trajectories} illustrate similar features that can be seen for the most massive part of the entire population $L$. For the close in planets we recover similar tracks to the brown group in Fig. \ref{fig:track_groups}. The outer tracks show similar tracks to the purple, violet and blue group although with enhanced growth, which results from the massive disk of the system. We can clearly see that with the inclusion of fragmentation we are able to form icy planets from initial embryos that are well within the initial ice line which is not the case without. This is due to the drift of fragments and the movement of the ice line during the simulation. {As the Ice line moves towards the star during the evolution of the gas disk \citep{Burn_2019}, icy fragments drift past initial location of the ice line enriching the interior embryos in volatiles. This does not happen for the planetesimals as their drift timescales are significantly longer meaning the icy planetesimals do not migrate far past the initial ice line.} Additionally we can see a distinct difference between the formation pathways of icy and rocky planets in these heavier disks where the growth of icy planets is greatly enhanced because of the drift pileup at the ice line caused by the transition in drift speed due to the change in bulk density of solids (by a factor of $\sim 3$). This can be seen by the sharp transition between the furthest initial embryo that forms a rocky planet and the closest icy planet forming one that differ in final mass by a factor of $\approx 5$ in mass {(marked with a black x in Fig. \ref{fig:trajectories})}. This also links to the gap of planets around the earths location discussed before seemingly being an effect of the sharp transition between icy and rocky planets. Additionally the distribution of the rocky planets also changes significantly leading to the formation of more massive planets forming from embryos starting close to the inner disk edge do to the increased access of accretable solids when compared to the non fragmenting case. So we can clearly see that the addition of fragmentation affects the formation pathways of planets very differently depending on the starting location of the embryos.

\begin{figure*}
    \centering
    \includegraphics[width = 0.8\hsize]{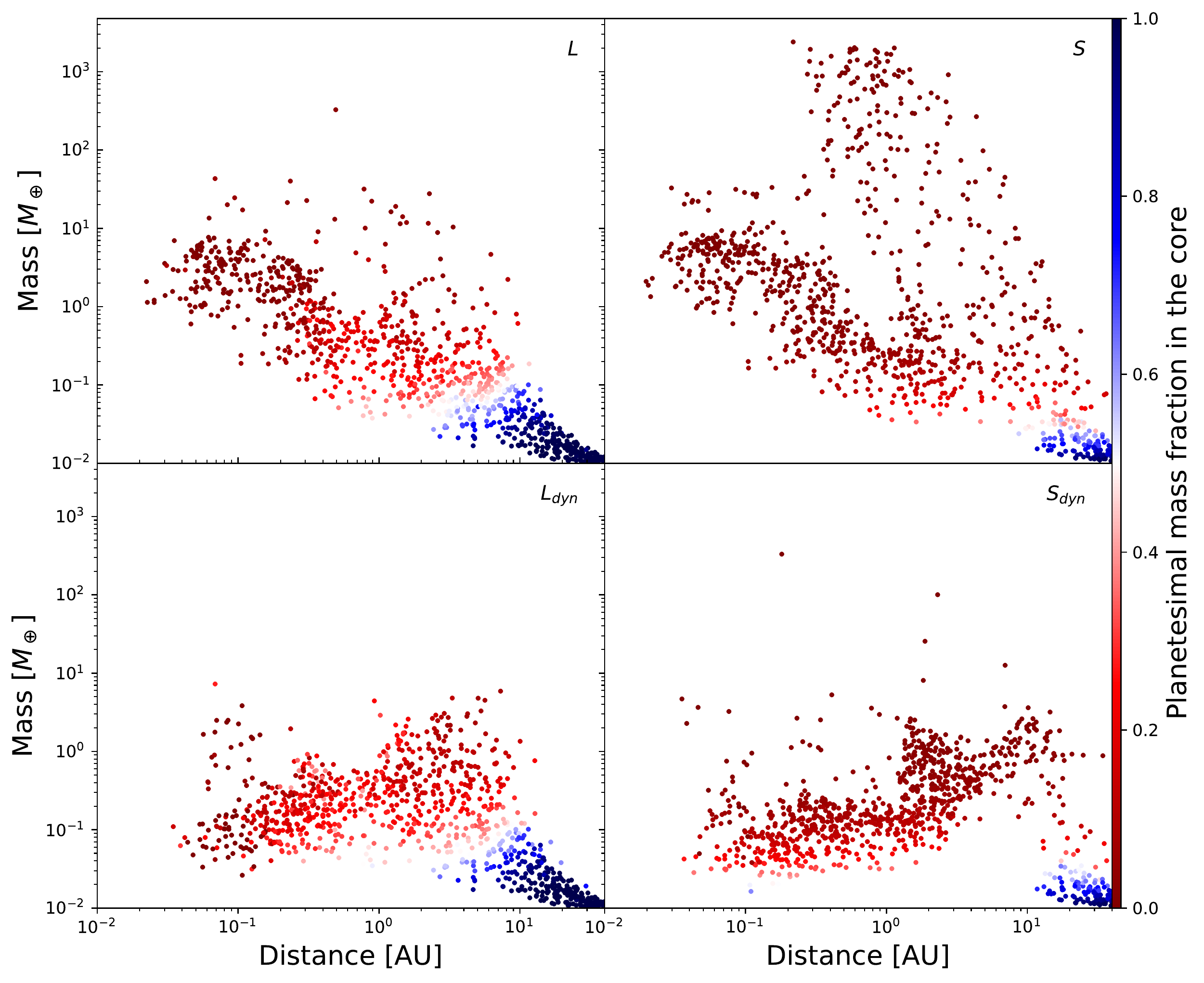}
    \caption{The distance Mass diagram for the populations with a single embryo ($100km$ (\textit{left}) and  $1km$ (\textit{right})). Where the colour refers to the mass fraction of planetesimals that were accreted onto the core}
    \label{fig:fragment_fraction}
\end{figure*}

The core mass budget in terms of fragments and planetesimals is an interesting statistic to look at as it allows us to directly see which is the most dominant mode of accretion i.e. how much mass was accreted in the form of fragments or planetesimals for the different types of planets. Additionally it lets us see the direct impact of fragmentation has on the formation of different types of planets. This is depicted in Fig. \ref{fig:fragment_fraction}.

A general trend for all populations is that in the far outer disk (from $\sim 10 AU$ outwards) fragmentation seems to have less of an effect on the accretion mode of planets as the embryos in these regions accrete negligible amounts of fragments. This can be explained by the low surface densities and long collision timescales among planetesimals leading to little fragments being produced and explains why the fragmenting populations are very similar to their non fragmenting counterparts beyond $\sim10-20 AU$. However this region is noticeably pushed outwards for the smaller planetesimals because due to their smaller size they have shorter collisional timescales. As a general trend in the populations with fixed fragment size, figure \ref{fig:fragment_fraction} shows that the closer we get to the star and the heavier the final planet is, the higher its mass fraction of accreted fragments is. When enabling the dynamic calculation of the fragment size we can a see similar trend with the caveat that we get low mass planets in the inner disk that accrete significantly less fragments and have a lower final mass of $\approx 10^{-1}M_\oplus$. This shows that the inner disk gets depleted of fragments when we consider these smaller fragments that have very short drift timescales which halts the growth of these embryos. The right column of Fig. \ref{fig:fragment_fraction} shows that more fragments are being accreted on the planets when compared to their $100 km$ counterpart which is true for both setups. This can be explained by the fact that the smaller planetesimal are much easier to fragment due to their lower material strength. In the end for the $1km$ the planets almost exclusively consist of fragments which implies that the initial $1km$ sized are very weak with regards to the collisions with them selves. 

\begin{figure*}
    \centering
    \includegraphics[width=0.8\hsize]{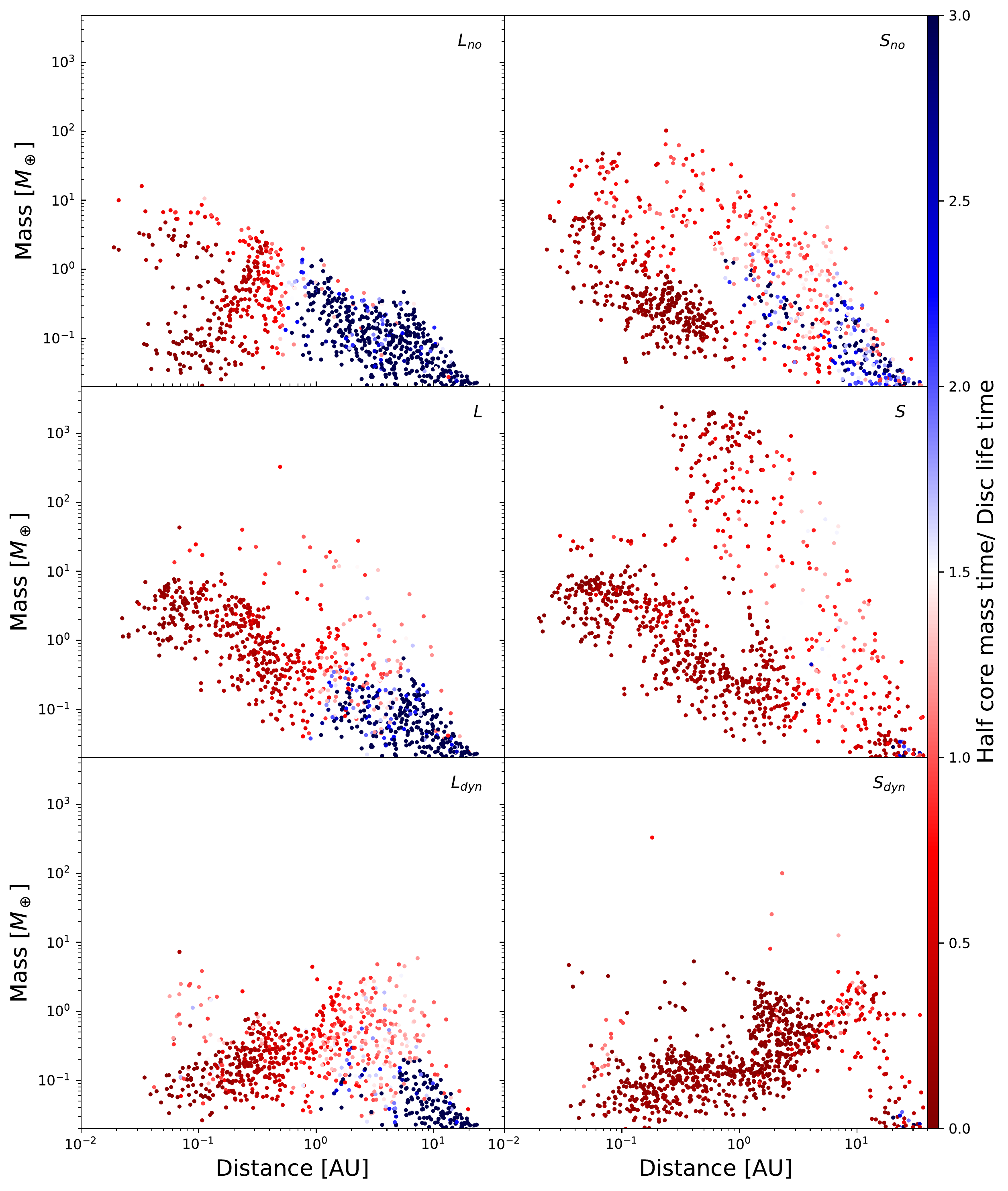}
    \caption{The mass distance diagram for the single embryo populations. The colour refers to the ratio of the time in which the embryos accrete half their final mass compared to the gas disk lifetime.}
    \label{fig:formation_time}
\end{figure*}

An additional important factor for planet formation is the timescale on which the planets form. This can be tracked by comparing the formation time of the core versus the lifetime of the gas disk. This largely dictates the amount of gas that can be accreted onto the planet for the heavier cores. This is especially important for the formation of giant planets as they have to accrete their gaseous envelop while the gas disk is still massive enough. To illustrate this we plot the time in which the core grows by half of its final accreted core mass versus the disk lifetime in Fig. \ref{fig:formation_time}.

Looking at the left column of Fig. \ref{fig:formation_time} for the $100km$ planetesimals we see a few key features: When we compare the formation times in the inner disk we can see that planets at the inner disk edge with $\sim 3-5 M_\oplus$ have shorter formation times when considering fragmentation compared to the ones without. Specifically we can see this quite clearly for the more massive planets ($M_p>M_\oplus$). An other interesting feature is the added late accretion for the planets further out (between $1-6AU$ and masses between $0.5-5M_\oplus$) which all have formation timescales around the disk lifetime and slightly above. The planets forming further out in the disk have formation times longer than the disk lifetime, however the fragmenting population $L$ displays a significant reduction in formation times for these planets. Looking at the the dynamic fragment size ($L_{dyn}$) we see additional speedup for the outer planets.

\begin{figure}
    \centering
    \includegraphics[width=\hsize]{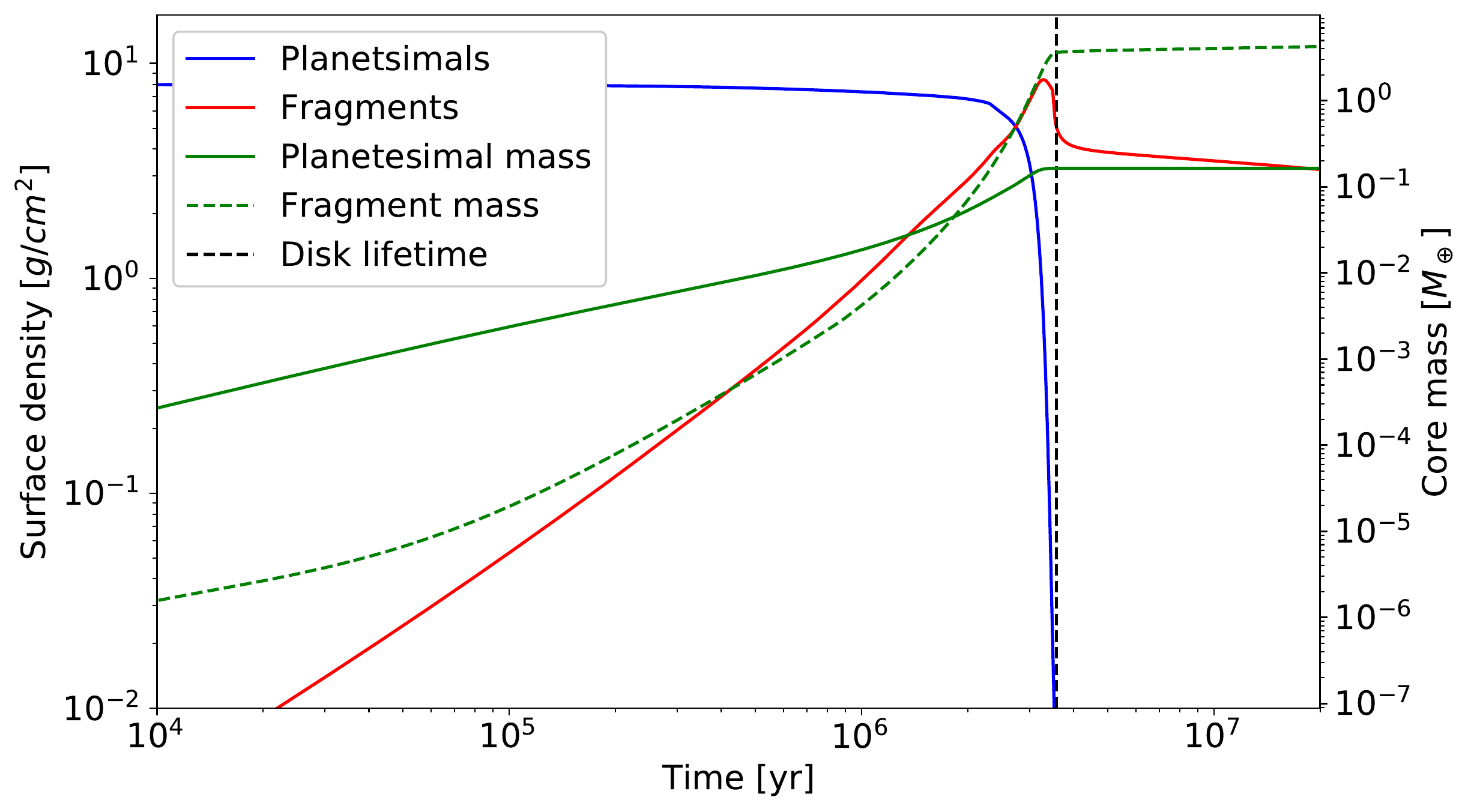}
    \caption{Core growth (\textit{green} right y-axis) for an example outer system (995 of $L$) with the surface densities (left y-axis) of planetesimals (\textit{blue}) and fragments(\textit{red}) and the time of the dispersal of the gas disk (\textit{black})}
    \label{fig:outer}
\end{figure}

For the $1km$ sized planetesimals we can see a significant reduction of the growth timescale when adding fragments of a fixed size which is in line with what we see for the $100km$ planetesimals. However the formation times get shortened significantly more so much that for population $S$ virtually all planets form within the disk lifetime. For population $S$ we can also see that the Giant planets have shorter formation times than the intermediate mass planets with $\sim 1-100 M_\oplus$. With the introduction of fragmentation we can observe an increase in occurrence rate of planets that grow to around $\sim 1 M_\oplus$ for population $L$ and between $ 1 -100  M_\oplus$ for $S$ beyond the ice line on the timescale of the disk lifetime. In order to illustrate what happens for these systems we show the growth track along with surface densities of fragments and planetesimals around an example planet in Fig. \ref{fig:outer}.

As we can see for this type of forming planet the generation of fragments occurs just before the dispersal of the gas disk. This means that the enhanced growth rate provided by the fragments only contributes during the later stages of planet formation. The fragments get generated during the end of the disk lifetime because with the removal of the disk the eccentricity damping from gas drag weakens and relative velocities are increased, leading to higher fragmentation rates. The generated fragments have much lower $e$ and $i$ than the planetesimals as long as the disk is still present and take some time to be stirred up as the disk vanishes. Which means they are accreted a lot faster. This leads to a significant accretion boost right around the time of disk dispersal that we can see in many simulations where the embryo is outside of $\sim3 AU$. This is an interesting feature as it implies that not only the stirring by the planet can lead to the generation fragments but the dispersal of the gas disk plays an important roll in the collisional evolution. This {enhanced} growth around the disk lifetime also has implications for the enrichment of heavy element in the envelops of these planets \cite{Shibata_2022} as we expect a significant amount of the planetesimals {accreted at later times when the planet has already accreted gas} to be deposited in the envelop.

\begin{figure}
    \centering
    \includegraphics[width=\hsize]{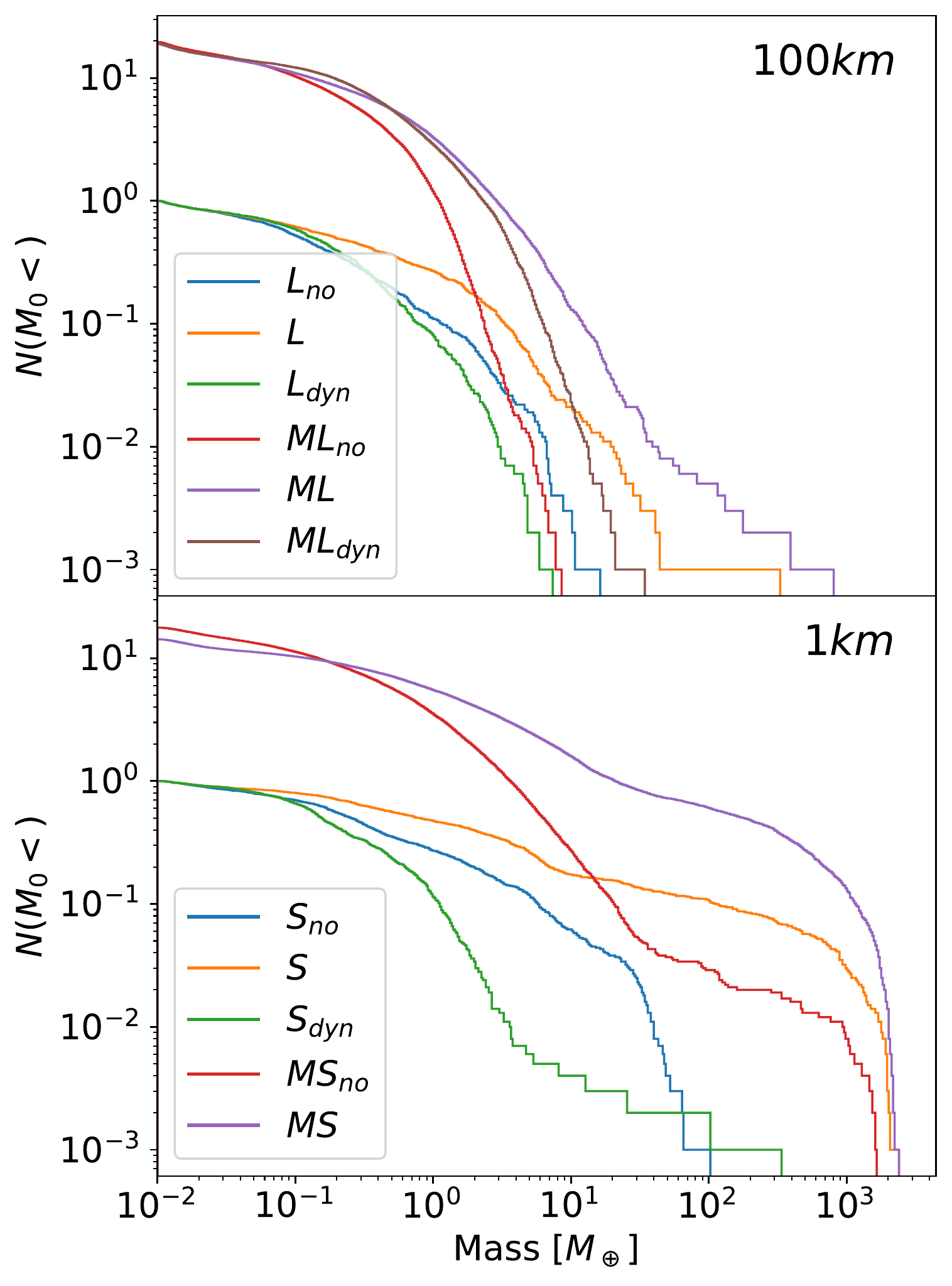}
    \caption{The reversed cumulative planetary mass function for the populations. It is normalised to the number of systems for all populations}
    \label{fig:cummulative_mass_function}
\end{figure}

An other important quantity to compare the planet populations is the mass distribution of the formed planets which is described by the planetary mass function (PMF). We show the PMF as a reversed cumulative distribution function for all our populations (including the multi embryo ones which will be discussed later) which can be seen in Fig. \ref{fig:cummulative_mass_function}.

As we can clearly see the addition of fragmentation leads to a significantly more massive population of planets when we consider a fixed size for the generated fragments which is true for both the $100km$ and the $1km$ planetesimals. However when we consider the dynamical size calculation of the fragments the picture changes quite drastically and we get an adverse effect across all mass ranges. The runs with smaller planetesimals as expected yields enhanced growth across the spectrum when compared to their $100km$ counterparts as the smaller planetesimals are more easily accreted. An interesting feature of population $S$ when compared to previous works \citep{NGPPS_2} is that the occurrence rate gap of planets around $100M_\oplus$ is much less pronounced which may be due to the late growing outer planets discussed before as they fit right in the mass range of the gap.

\begin{figure*}
    \centering
    \includegraphics[width = \hsize]{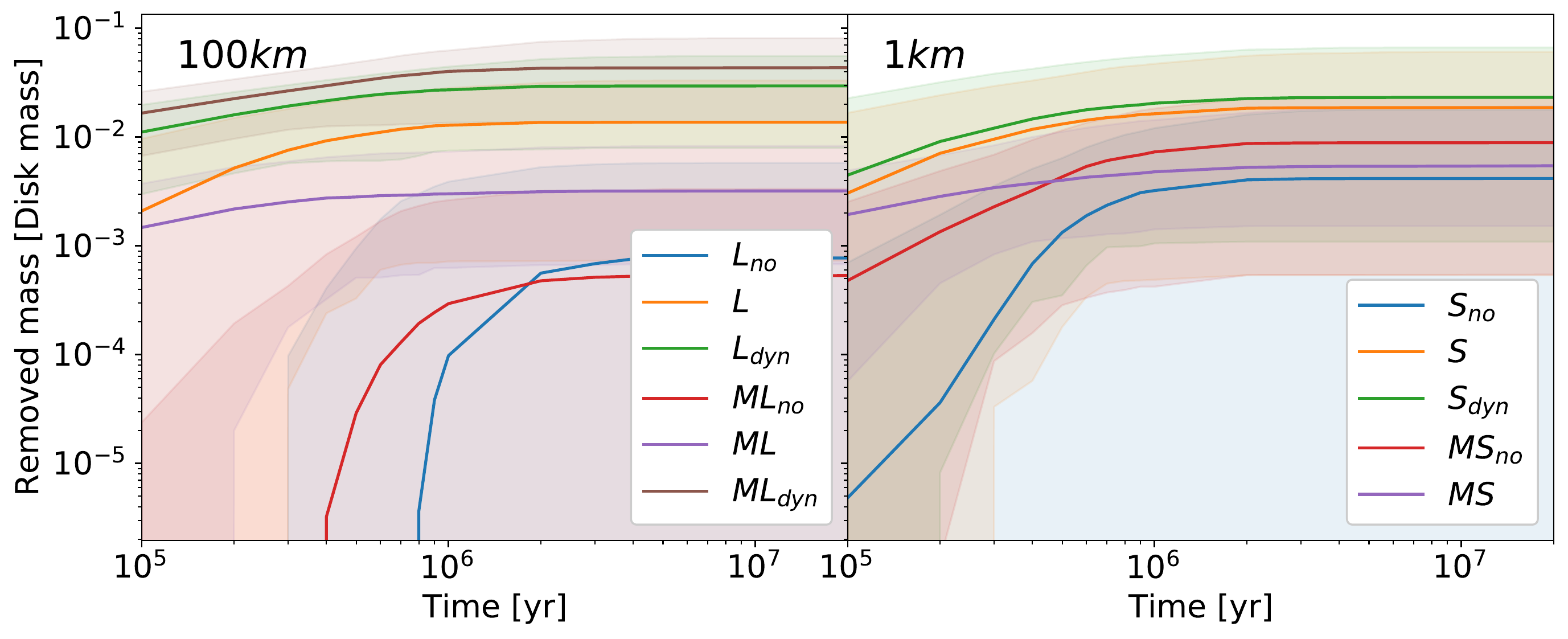}
    \caption{The median mass removed form all the disks in each population at different times along with the 10-90 percentile interval(shaded) for the bigger planetesimals (\textit{left}) and the smaller ones (\textit{right})}
    \label{fig:m_remove}
\end{figure*}

With the introduction of the radial drift of solids, a significant amount of mass {is transported} to the global pressure maximum at the inner disk edge which leads to a pileup of solids as discussed in section \ref{sec:soliddisk}. This happens because in our disk models this trapping of solids happens outside of the sublimation line for refractories. In order to approximately treat the not resolved complex physics at the inner disk edge, we limit the solid surface density to the gas surface density above a limit of $200 g/cm^2$. In Fig. \ref{fig:m_remove} we display the mass removed by this limiting of surface density for the different populations during the formation stage of the planet formation. We do this by plotting the median mass removed along with the 10-90th percentile interval for all systems. This serves as a good measure for the mass transport in the inner disk as it allows us to see how much additional material is transported to the inner disk edge.

As we can clearly see in Fig. \ref{fig:m_remove} the addition of fragmentation leads to much earlier mass removal i.e. increased mass transport to the inner disk. This is enhanced even further when we consider the dynamical calculation of the fragment size which is expected as the smaller fragments have shorter drift timescales then bigger ones. Additionally for the $1km$ planetesimals we observe little difference (except populations $S_{no}$ and $L_{no}$) when compared to their $100km$ counterparts owing to the fact the the mass transport is dominated by the fragments and not the planetesimals. Additionally we can see in Fig. \ref{fig:m_remove} that this affects the populations without planetesimal fragmentation mostly on the timescale of lifetimes of the gas disks which limits its imprint on the forming planets especially as the formation timescales are short in the inner disk (see Fig. \ref{fig:formation_time}). However for the populations with planetesimal fragmentation we remove this mass quite early meaning we experience a lot of drift in the early stages depleting the inner disk of solid material. This is one of the main explanations why growth is halted for embryos in the inner disk when we consider the dynamical size of the fragments. 

In order to investigate the radial extent of the mass removal in the inner disk and its impact on the forming planets we ran an additional population with the same initial conditions as population $L_{no}$ but {without} the limiting of the surface density. We chose this population because the surface density profile steepens with shorter drift timescales so population $L_{no}$ is affected out to the largest distance from the star when compared with the other populations. This makes it the most conservative choice showing an upper limit for the radial influence of this treatment. When we compare the final masses of the same planets forming in the two populations we see that the differences are negligible outside of 0.4 AU with the maximal relative difference in mass being $7*10^{-3}$ and an average of $3.5*10^{-4}$. However we should be aware that it impacts the planets forming at the inner disk edge where we see an average mass deviation of $7\%$ inside $0.4AU$. This means we have to be careful when interpreting the results of our simulations with forming planets close to the inner disk edge. However we know that we underestimate the masses of these planets so the resulting final planet masses serve as a lower limit.
 
\subsection{multi embryo population}\label{sec:multi}   
\begin{figure*}
    \begin{center}
    \includegraphics[width = 0.5\hsize]{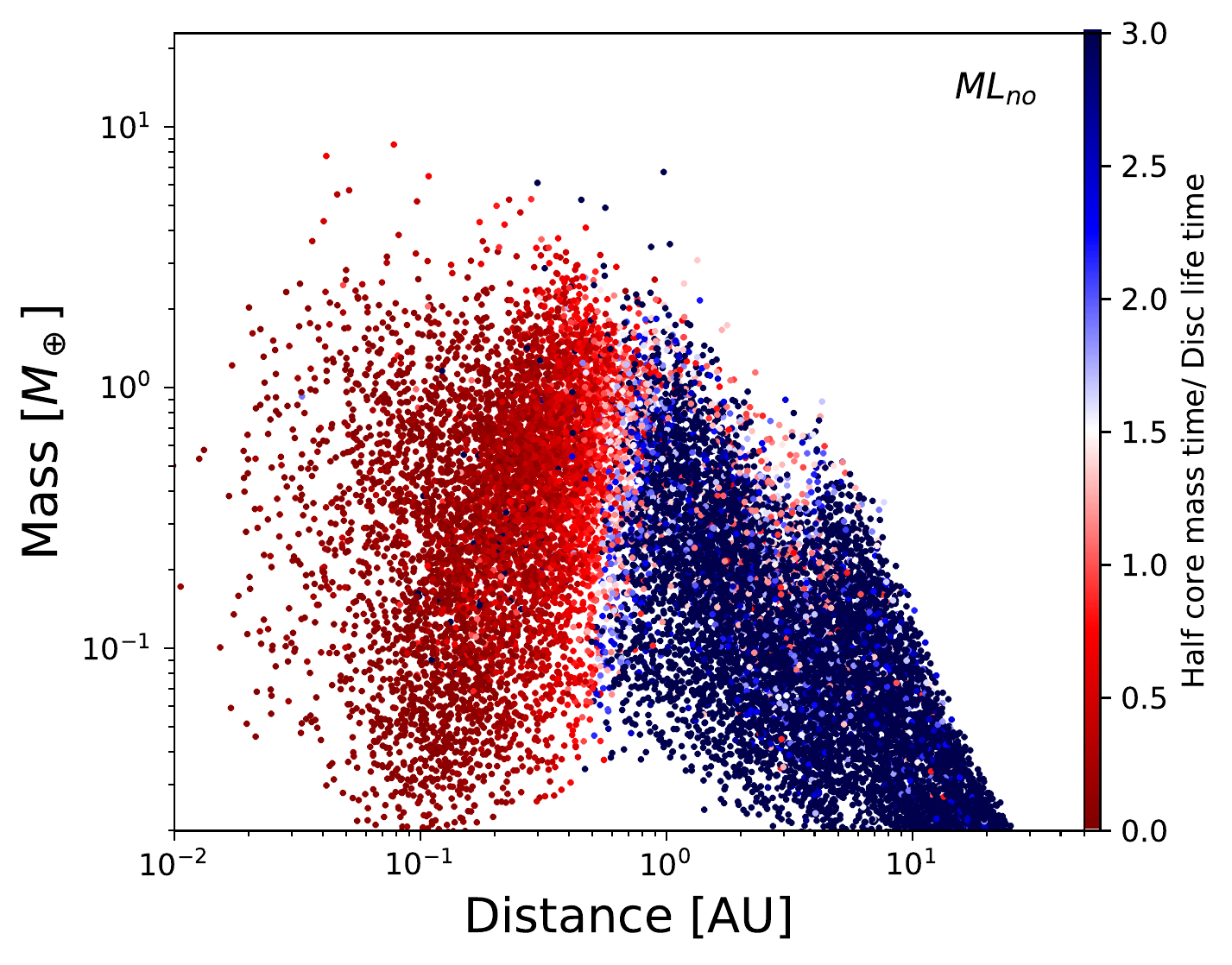}
    \end{center}
    \includegraphics[width = 0.5\hsize]{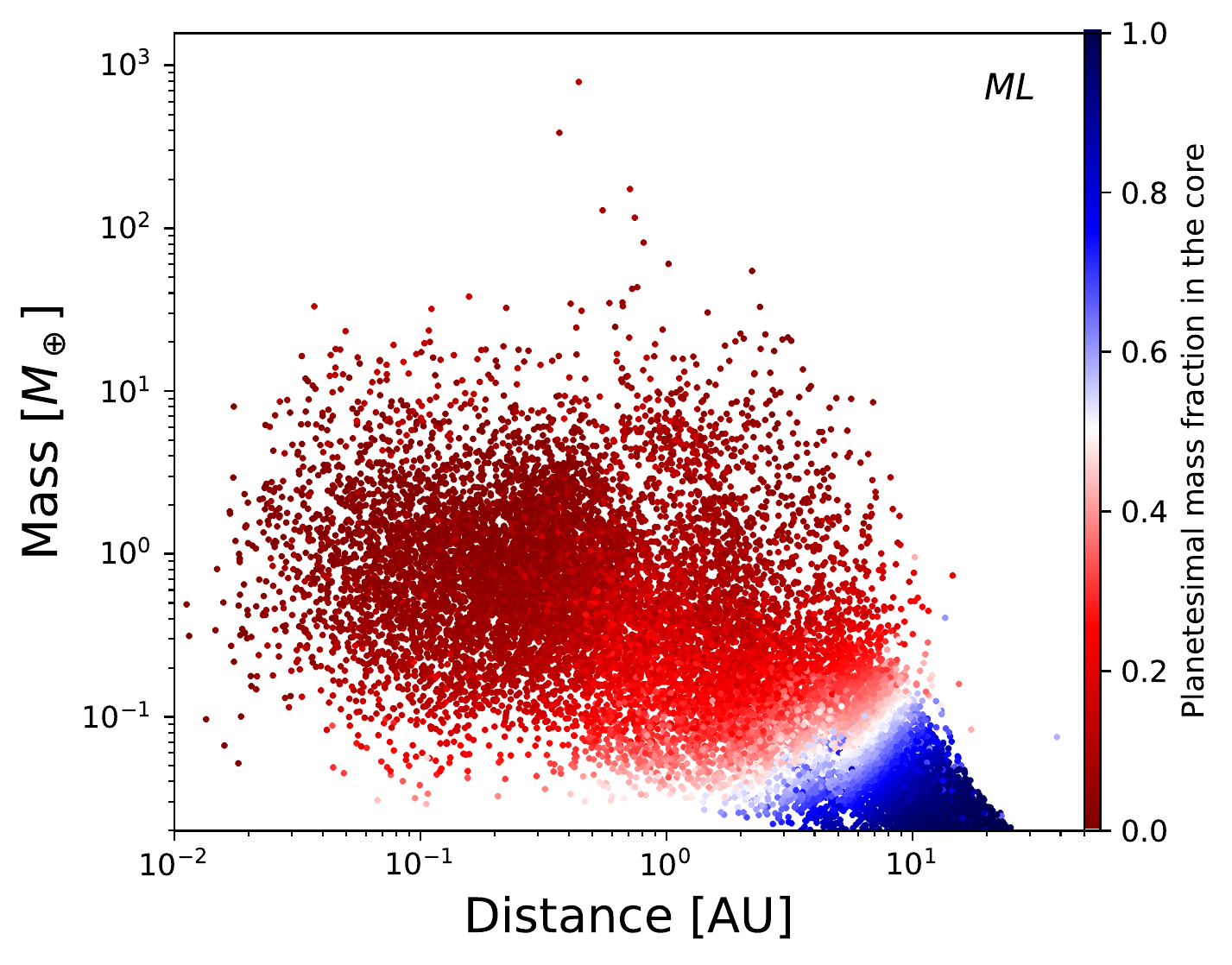}
    \includegraphics[width = 0.5\hsize]{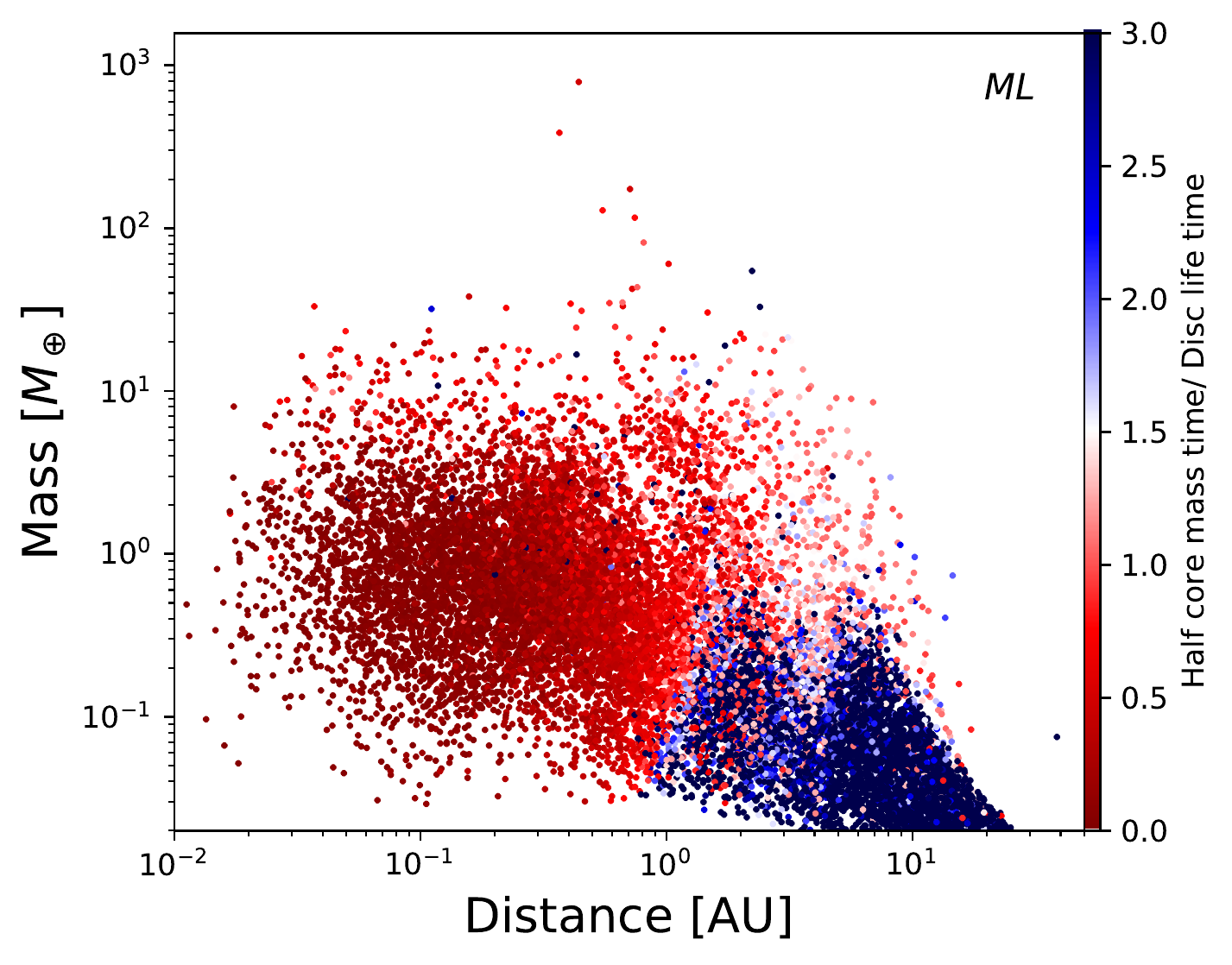}
    \includegraphics[width = 0.5\hsize]{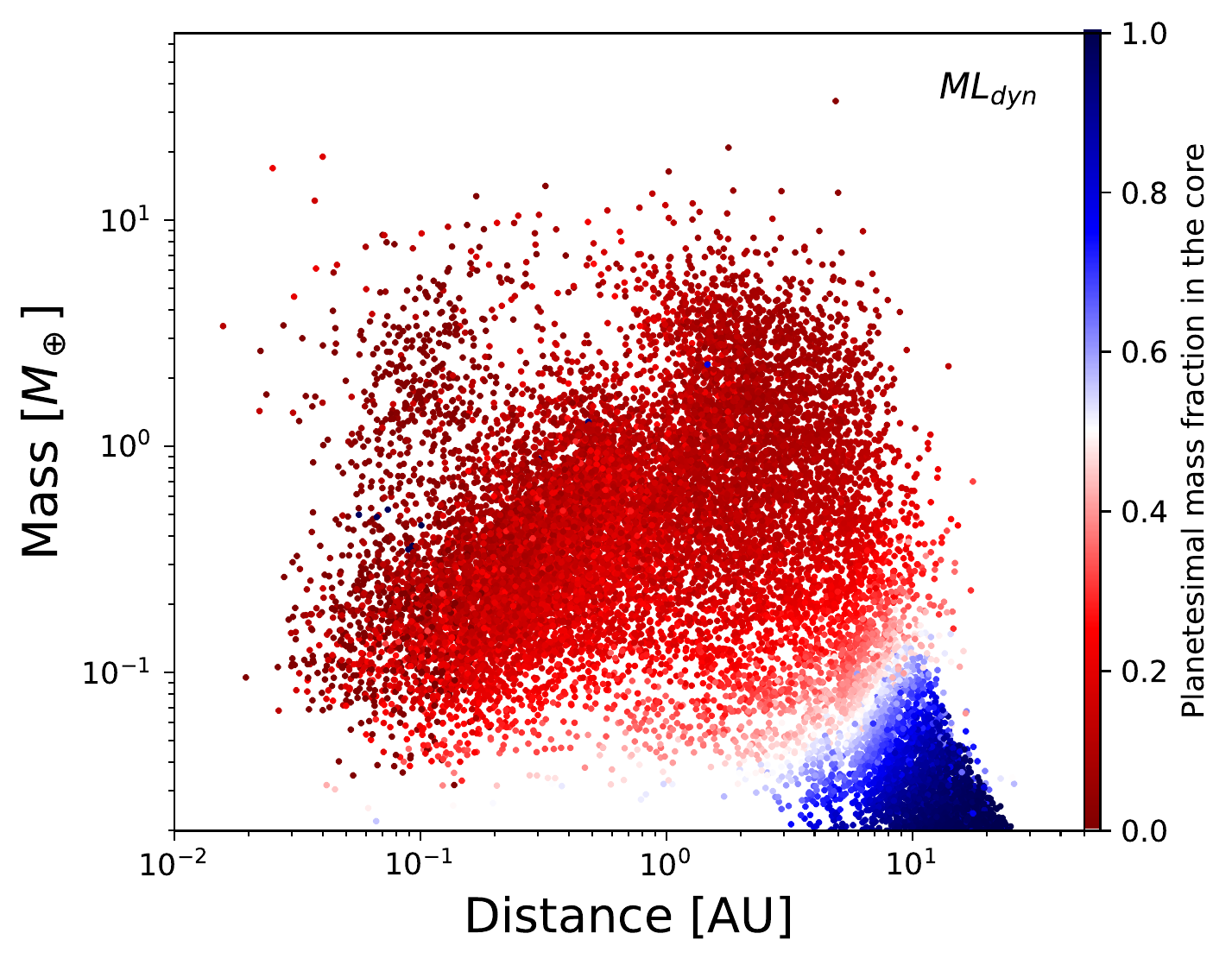}
    \includegraphics[width = 0.5\hsize]{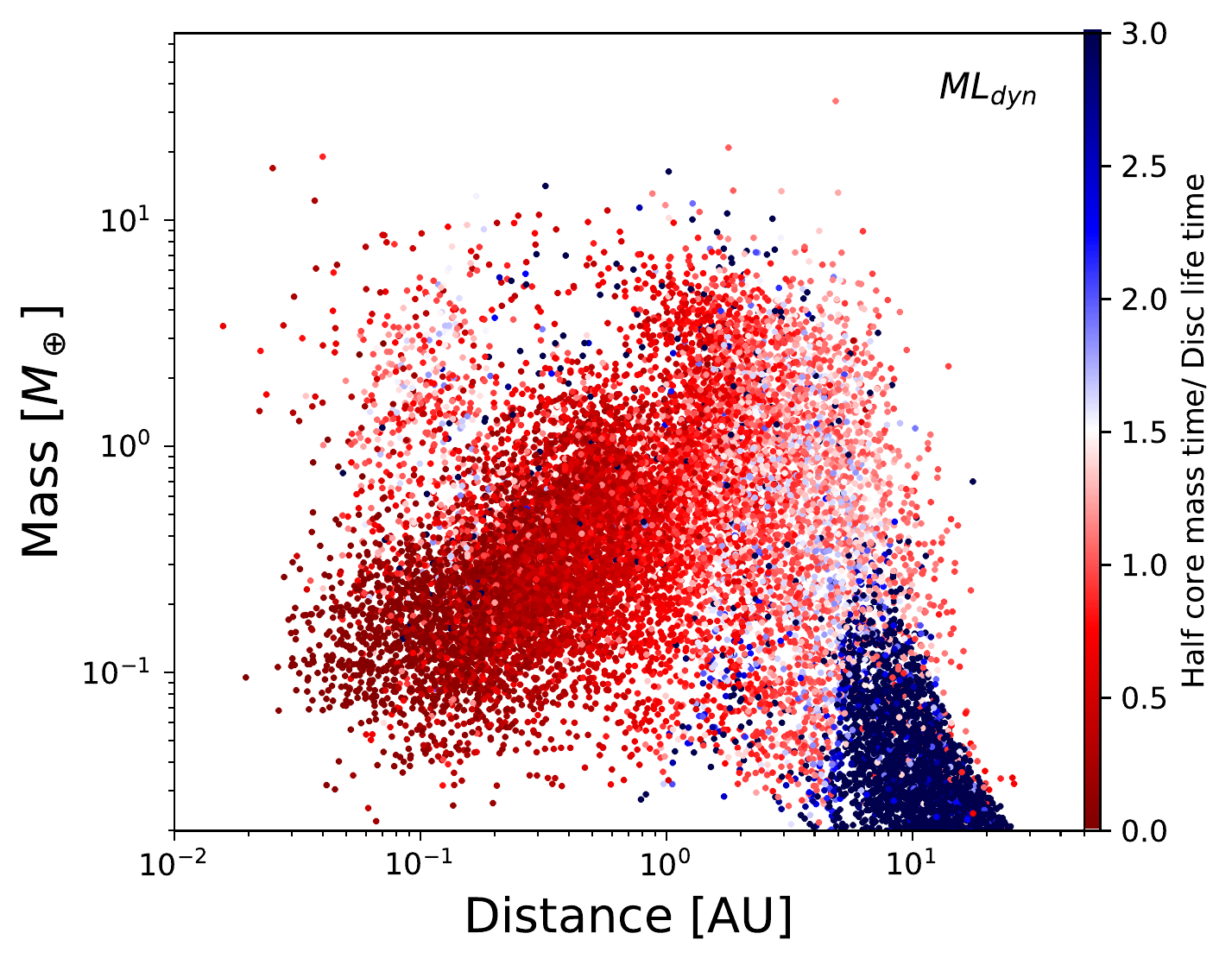}
        \caption{The distance mass diagram analogous to Figs. \ref{fig:fragment_fraction} (\textit{left}) and \ref{fig:formation_time} (\textit{right}) for populations $ML_{no}$ (\textit{top}), $ML$ (\textit{middle}) and population $ML_{dyn}$ (\textit{bottom}) with 20 embryos per disk where only the surviving planets are included. Note that the mass limits of the figures vary for the different populations}
    \label{fig:Multi_smaL}
\end{figure*}

\begin{figure*}
    \begin{center}
        \includegraphics[width = 0.5\hsize]{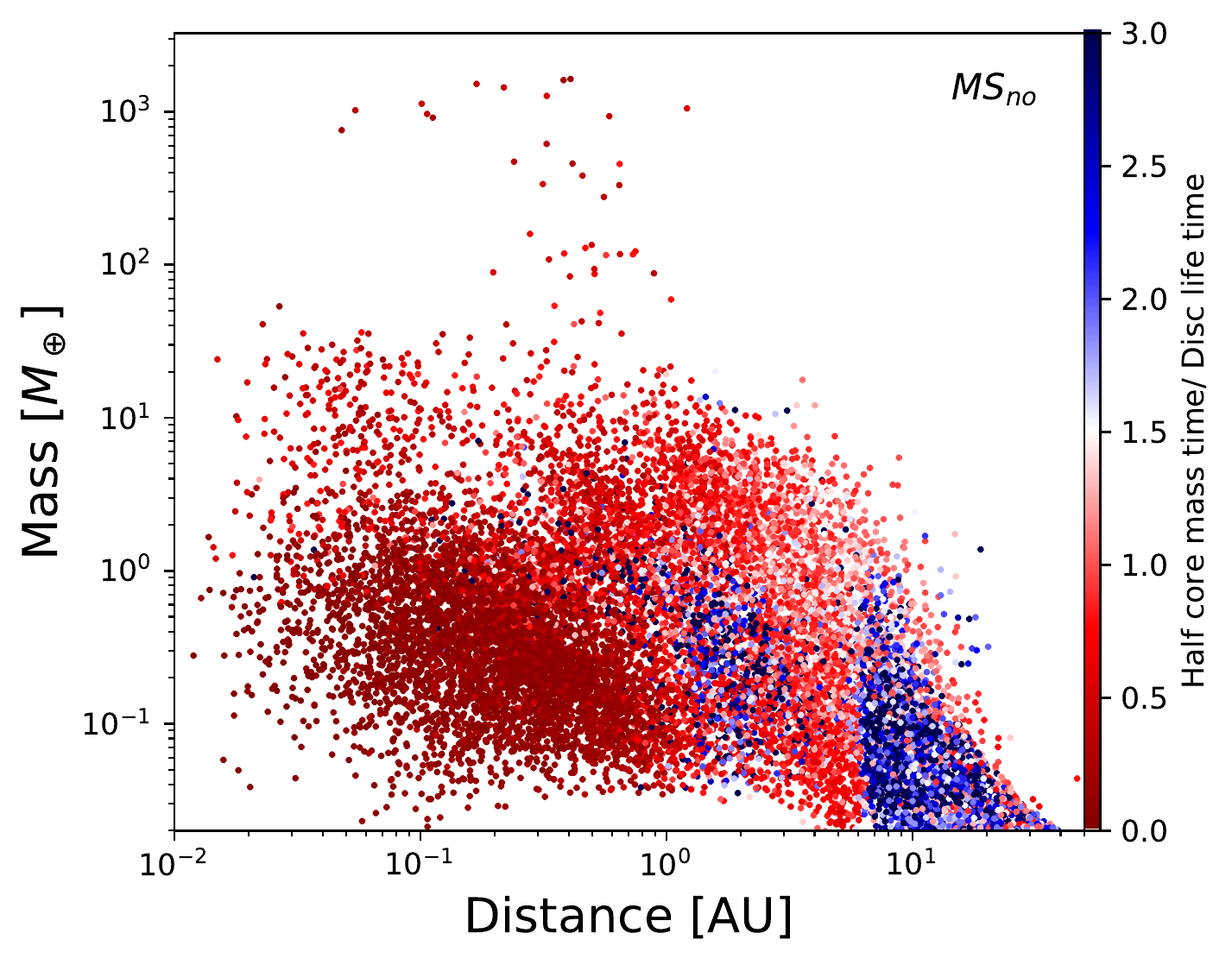}
    \end{center}
    \includegraphics[width = 0.5\hsize]{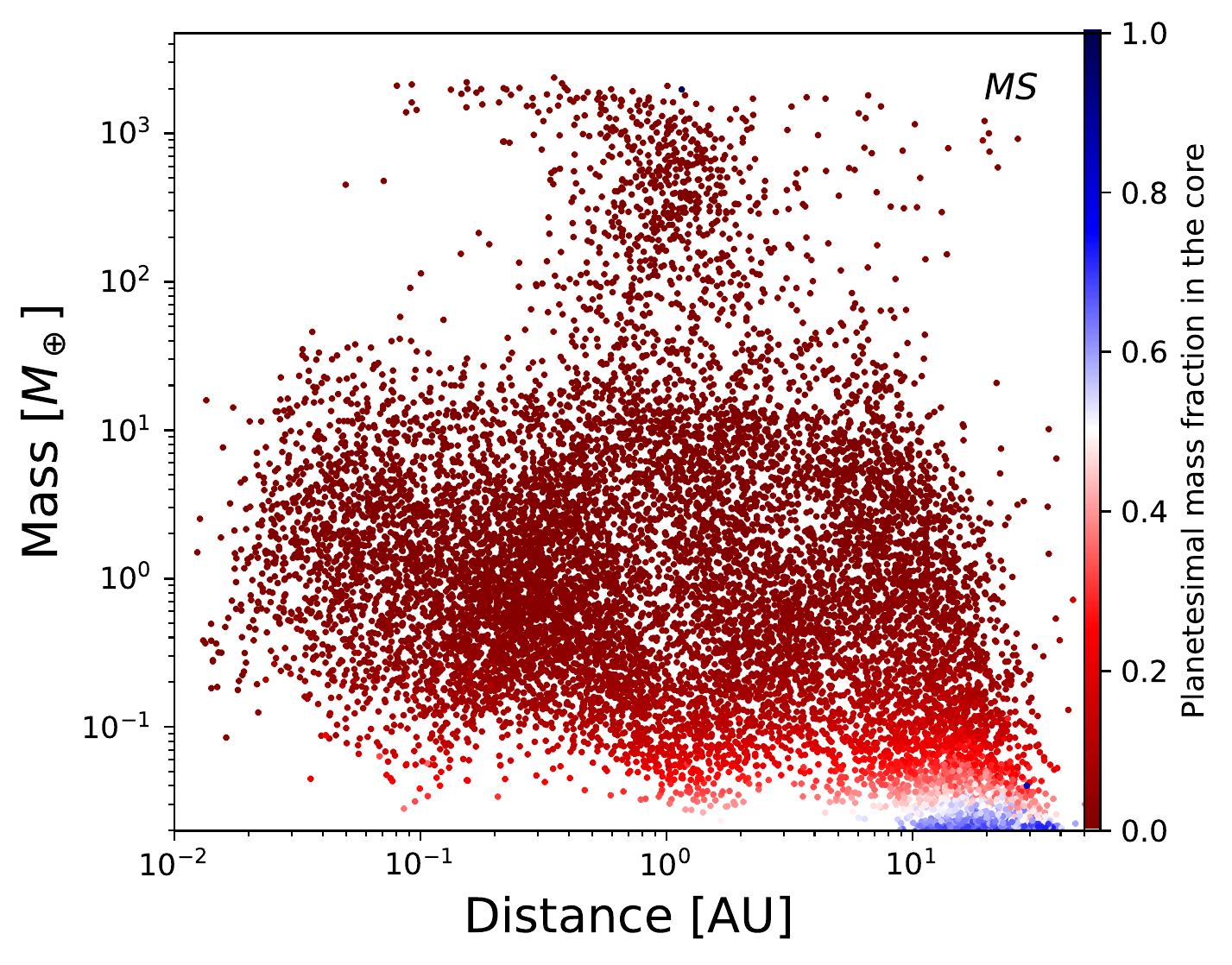}
    \includegraphics[width = 0.5\hsize]{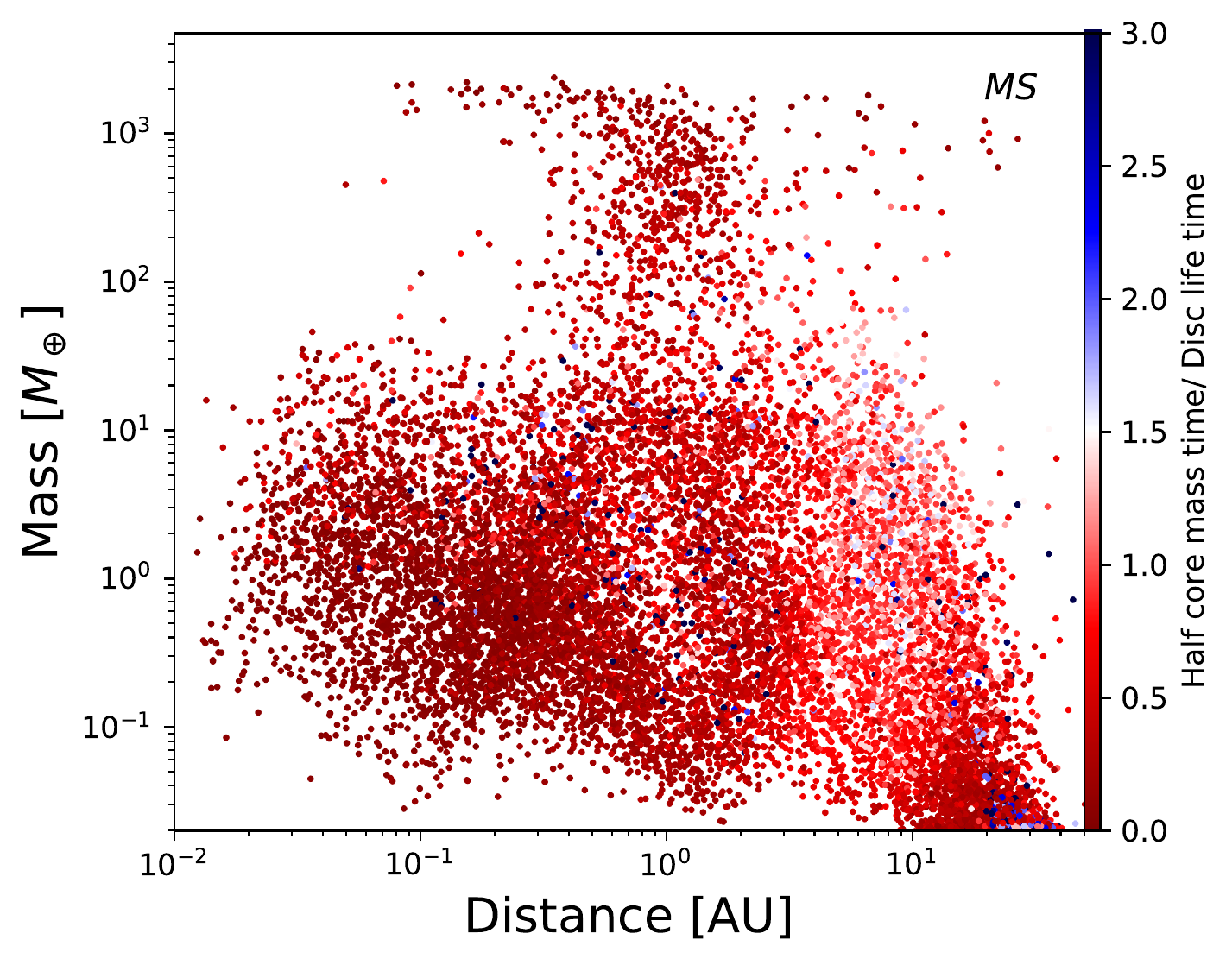}
    \caption{The distance mass diagram analogous to Figs. \ref{fig:fragment_fraction} (\textit{left}) and \ref{fig:formation_time} (\textit{right}) for populations $MS_{no}$ (\textit{top}), $MS$ (\textit{bottom}) with 20 embryos per disk where only the surviving planets are included.}
    \label{fig:Multi_smaS}
\end{figure*}

It is also interesting to investigate the formation of multiple planets in the same disk because fragmentation and the other added processes open up further possibilities of interaction between planets. For example the accretion of the inwards drifting fragments generated by another planet {further out}. To investigate this we run a population with the same parameters as $L_{no}$, $L$, $L_{dyn}$, $S_{no}$ and $S$ (named $MX$ where $X$ is the single embryo name) that include $20$ initial embryos per system to see if there are any emerging imprints on the forming planets left by the interplay between the presence of multiple embryos per disk and planetesimal fragmentation. To investigate this {we show in Figs. \ref{fig:Multi_smaL} and \ref{fig:Multi_smaS} the same quantities as in Fig. \ref{fig:fragment_fraction} and \ref{fig:formation_time} for the multi planet populations. Note that we did not plot the planetesimal mass fraction in the core for populations $ML_{no}$ and $MS_{no}$ as it adds no new information (cores are made up only of planetesimals).} 

When we compare the final masses and positions of the planets in the reference simulation ($ML_{no}$) to its fragmenting counterpart ($ML$) we recover very similar effects of fragmentation on the population as we have in the single embryo case. Namely, the features we observe are, significantly enhanced growth beyond the ice line including the increased occurrence rate of massive planets with formation times around the dispersal of the gas disk outside of the ice line. Additionally we also see the promoted growth at the inner disk edge where we don't have many planets in the $10^{-2}-10^{-1}M_\oplus$ mass range anymore when considering fragmentation. When considering the dynamic treatment of the fragment size ($ML_{dyn}$), we can see the significantly lower planet masses inside of $1AU$ when compared to population $ML$ along with the enhanced growth in the outer disk which are both features we recovered from comparing the respective singe embryo populations. For the smaller planetesimals When looking at the population $MS_{no}$ we see that with the addition of multiple embryos giant planets are forming in the disks which is not the case for $S_{no}$ as giant impacts between the embryos lead to enhanced growth. For the population $MS$ one can see that the gas giants have much wider distribution of semi major axes when compared to their single embryo counter parts which is due to the n-body interactions between the growing embryos. By comparing the two ($MS_{no}$ and $MS$) we can see that the addition of fragmentation clearly enhances growth in the outer disk shortening formation timescales beyond the ice line.

The fragment fractions displayed in the left column of Figs. \ref{fig:Multi_smaL} and \ref{fig:Multi_smaS} take into account the composition of the colliders for embryo mergers i.e. the planetesimal fraction of the merged core is computed consistently from its colliders. When we compare the fragment fraction of the multi embryo populations with their respective single embryo counterpart we can see the same general trends, higher fragment fractions for closer in and heavier planets. When we look at the simulations with bigger planetesimals ($ML$) the picture remains largely the same, however due to the scatter introduced by the N-body interactions we have a less homogeneous picture, for example we can see quite a few planets containing little fragments in the inner system. Additionally for the $ML_{dyn}$ we see that most low mass planets inside of 1AU have higher planetesimal mass fractions in the core compared with the fixed size treatment. For the $1km$ sized planetesimals we once again see that apart from the very low mass planets all the mass gets accreted in the form of fragments. The right column of Fig. \ref{fig:Multi_smaL} and \ref{fig:Multi_smaS} depicting the formation times also generally shows the same trends for the as discussed in Section \ref{sec:single}. In these plots we can nicely see the random scatter from the N-body interactions with the planets with very long formation times (dark blue) which are distributed throughout the populations including fragmentation. These planets grow from giant impacts and therefore are neither restricted by the lifetime of the gas disk nor can they be present in the single embryo case. Remaining planets display the same distribution of formation times as their single embryo counterpart.

The PMF for the multi embryo populations are already shown in Fig. \ref{fig:cummulative_mass_function}. The y-intersect of the PMF is at less then the initial number of embryos due to the fact that only the surviving planets at 5 Gyr are being counted for the statistic which means all the planets lost to collisions and the star are omitted. As for the single embryo case the addition of fragmentation (i.e. comparing $ML_{no}$ and $ML$) with a fixed fragment size leads to a significantly more massive planets. However, opposed to the single embryo case the dynamic fragments also produce more massive planets which is due to the large number of planets with masses between $1-10M_\oplus$ outside of $1AU$. For the 100km planetesimal populations $ML$ and $ML_{dyn}$ we see that only very few planets get lost to collisions which is due to their overall lower mass when compared to the simulations with $1km$ sized planetesimals. When we look at the population $MS_{no}$ we clearly see the imprint of the giant planets on the PMF which are not present in the single embryo counter part leading to a much higher maximal mass. For the smaller planetesimals with fragmentation ($MS$) we see very similar results when compared to their single embryo counter part however the lack of planets around $\sim 100 M_\oplus$ is less pronounced then expected but more pronounced then in the single embryo case. The addition of fragments of a fixed size still seem to enhance the growth of the planets significantly and also lead to more planets being lost (i.e. an increase in the embryo embryo collisions that occur). The maximal masses appear to be almost the same for the single and the multi embryo cases for the three simulations which is in line with the results from \cite{NGPPS_2} where the maximal mass shows only a slight dependency on the initial number of embryos.

When we look at the mass that was removed due to the limiting of the surface density in the inner disk displayed in Figure \ref{fig:m_remove} we see that there is less mass removed when we compare the multi embryo populations to the single embryo equivalents which is to be expected because the more embryos exist in the disk the higher the chance one starts in the inner disk and accretes some of the material before it is lost to the removal. Additionally the mass flux to the inner disk edge gets reduced {for a larger number of embryos as they accrete the fragments before they reach inner disk edge.} This trend does not hold when we compare $MS_{no}$ with $S_{no}$ which may be explained due to the added presence of giant planets which stir up the planetesimals leading to increased drift speeds \ref{eq:v_drift}.

In summary we can see that the presence of multiple planets does not seem to break the imprint planetesimal fragmentation has on planet formation, however there are still a few changes that arise due to the multi planet nature of these systems. This is in line with the results of \cite{Guilera_2014} that finds little change for two in situ forming planets in the same disk. However in this work we explored the possible effects of multiple forming {migrating} planets on many different systems and planet pairs leading to a more robust confirmation of the lack or little influence of the interplay. 
    
\subsection{Limitations of the model}
\label{sec:limitations}
\begin{figure}
    \centering
    \includegraphics[width=\hsize]{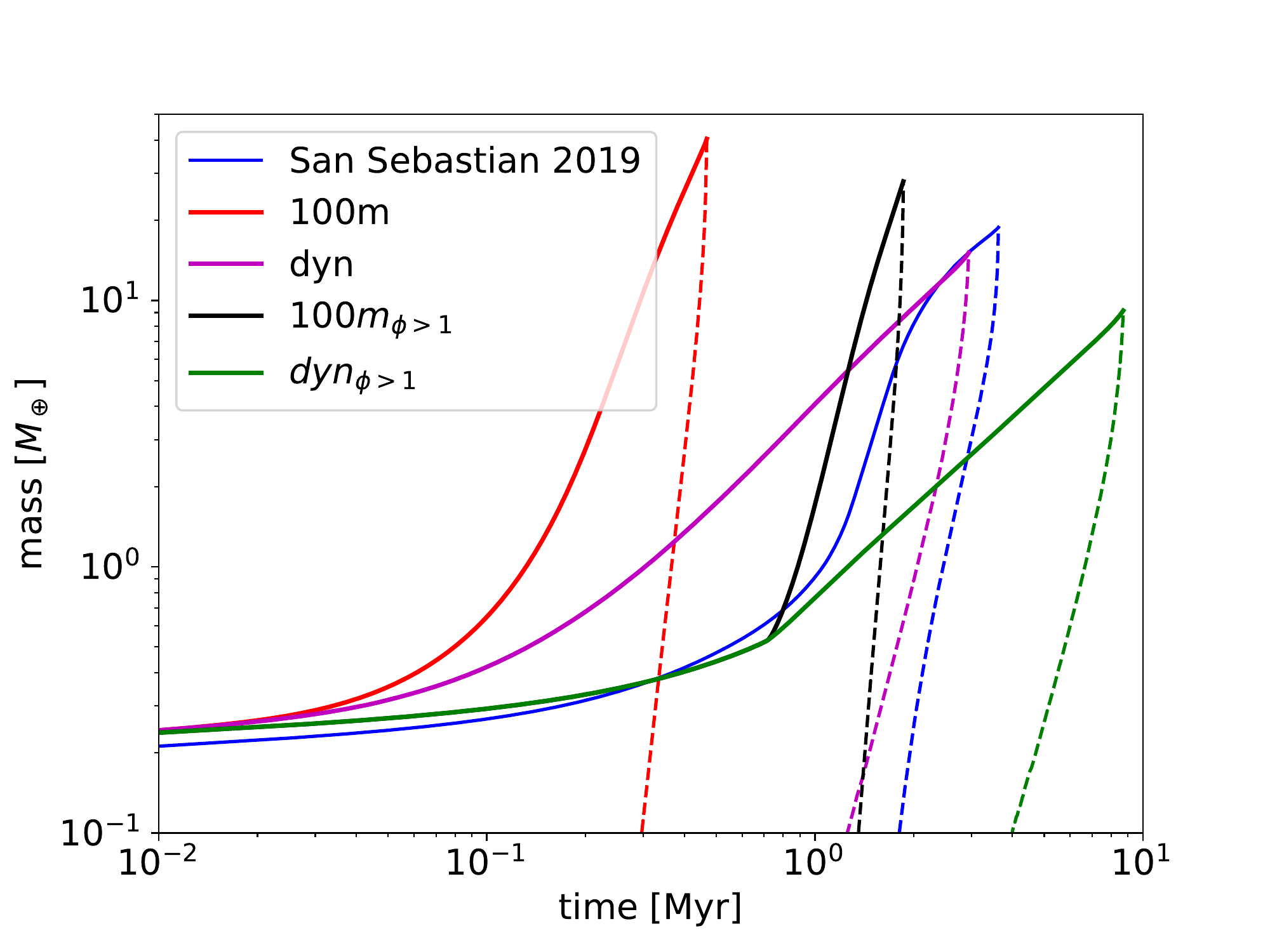}
    \caption{The growth tracks of an in situ forming planet at 5 AU in a ten times MMSN calculated with the model of \cite{San_Sebastian_2019} (blue) and our model with fixed size (red,black) and the dynamic size calculation (magenta,green) in the nominal case and when only consider collisions once $\phi >1$. Where the solid lines are core masses and the dashed ones correspond to the envelope masses.}
    \label{fig:compare_san}
\end{figure}

Our model describes the size distribution of solids with two bins as opposed to treating the entire size range of solids. We do this to reduce the computational cost of the simulations and to have a conceptually simpler model. This however means that we do not account the fact that the fragments created by collisions have a size distribution rather then a characteristic size. This becomes important for the collisions among fragments in the typical size range we consider ($\sim 100m$) as they become super catastrophic ($Q/Q_d^*\gg 1$) \citep{Guilera_2014} and lead to mass being transported to smaller sizes. This leads to a loss of accretable material (via drift or being reduced to dust sizes) which is ignored in our model. The only way this is considered at the moment is when considering the dynamical size change by assigning a smaller typical size to the fragments for which the collisions are not super destructive anymore. In order to investigate the effect of the single size treatment of fragments we can compare our results with the giant planet formation model of \citep{Guilera_2010,Guilera_2014,San_Sebastian_2019} that includes fragmentation and the full size distribution. We simulate the in-situ formation of an initially $0.05M_\oplus$ mass embryo in a 10 MMSN disk as described in the baseline case of \cite{San_Sebastian_2019}. The growth is calculated until the crossover mass is reached. To be consistent with their results we use the fragmentation energy description of \citep{Benz_1999} for basalt at $3km/s$ and assume a bulk density of $1.6g/cm^3$ for the solids. We run our model both with a fixed fragment size of $100m$ (red) and the dynamical size calculation (magenta). The resulting growth tracks of the embryo computed with our model and theirs (blue) can be seen in Fig. \ref{fig:compare_san}. Their resulting formation times are considerably different as our simulations display shorter formation times in such a massive disk. One reason for this is that they only consider the generation of fragments once the collisions among planetesimals reaches $\phi =  Q/Q_d^* >1$ (where $Q$ is the specific impact energy and $Q_d^*$ is the specific fragmentation energy) i.e. when collisions become fully fragmenting which is not the case in our model. To account for this we ran a second set of simulations with our model (black $100m$,green $dyn$) also adopting that model choice. With these additional changes we see that the formation time and crossover mass of the forming planet lies in between our different fragment size treatments which illustrates nicely the influence the size distribution has on the formation of planets.

Additionally we have approximate treatment of the inner disk because without the full size range we lack the information to apply more sophisticated models for the mass loss and solid evolution at the global pressure maximum \citep{Li_2021}. However previous studies have simply treated the inwards drifting material as direct mass loss which can account for a significant amount of their mass budget \citep{Chambers_2008}. Furthermore at the moment we only account for the sublimation of volatiles at the water ice line and do not add the sublimated material back to the gas disk. {However, this is reasonable as water ice makes up a major part of the volatile mass fraction leading to the most significant change in planetesimal properties}. We also assume local collision rates for the planetesimals i.e. we do not account for the fact that planetesimals with high eccentricities collide with planetesimals in neighbouring radial bins \citep{Morbidelli_2009,Guilera_2014}.

\section{Summary and conclusions}
\label{res}

In this work we investigated the influence of planetesimal fragmentation on the planet formation process by adding planetesimal fragmentation to the Bern model along with an updated description of the solid disk. Using the population synthesis approach we probed its impact on the formation on the different types of exoplanets for different planetesimal sizes and model parameters. The main results of this analysis are:
\begin{itemize}
    \item The addition of fragmentation does not allow for the formation of giant planets for $100km$ planetesimals. However it does promote the growth in the outer regions of the disk which is enough to enhance/enable the giant planet formation for smaller $1km$ sized planetesimals.
    
    \item For smaller fragments we can expect a hindering effect of planetesimal fragmentation on the planet formation process especially in the inner disk where the drift timescales are very short. For $100km$ planetesimals however they can still promote the growth of planets outside of a few AU both in single and in multiple embryo simulations.
    
    \item We find fragmentation promotes late growth beyond the ice line where we get a boost of accretion rate from fragments which are generated around the time of the gas disk dispersal as the damping by gas drag weakens.
    
    \item The simulations highlight the significance of how the inner disk edge/gas pressure maximum is treated for planet formation. In consequence our results in the inner disk (<$0.4AU$) are to be treated as preliminary due to the shortcomings of the model.
    
    \item The presence of multiple embryos in the disk does not significantly alter the imprint of fragmentation onto planet formation.  
\end{itemize}

As we see significant changes for the populations when considering fragmentation we will discuss it's implications on previous results obtained with the Bern model \citep{NGPPS_2,NGPPS1,Voelkel_2020}. In the works of \citep{Voelkel_2020} they also performed single embryo populations via the accretion of $100km$ planetesimals. However, they investigated the influence of different surface density slopes for the Planetesimals along with the dynamic formation of planetesimals. Due to the differing initial conditions in both works a quantitative comparison is not feasible so we will discuss the emerging features form the investigated added physics. The effects of a steeper density slope and the addition of fragmentation for the nominal case show similar imprints on the populations in the inner disk where we get significantly enhanced growth. However the results start to vary beyond the ice line where the steeper density profiles has little effect on the population where as fragmentation gives a significant boost to planetary growth as icy planetesimals fragment faster. Also with the consideration of the dynamic fragment size the picture changes quite a lot preventing most planets forming within the ice line to grow beyond $0.5 M_\oplus$. So the consideration of fragmentation could depending on the model chosen enhance/counteract the effects of the dynamic embryo creation has on the forming planets as found in the works of \citep{Voelkel_2020}. This would make it important to consider the effects in tandem in future works, especially since the consideration of both processes increase the self consistency of the solid disk and both have a non-negligible effects on the formation of planets.

There are several improvements and followup questions we want to explore in future works. The single size treatment of fragments is not ideal when considering a collisional outcome model with varying slope for the size distribution of fragments \cite{Guilera_2014}. This means a more elaborate description of the full size distribution of solids should be added incorporating a model for the dust evolution and planetesimal formation \citep{Guilera_2020,Voelkel_2020}. We showed the importance of this by comparing our results with the works of \citep{San_Sebastian_2019} leading to significant differences depending on the treatment of the fragment size. Additionally the impact of different descriptions of the fragmentation energy is also of interest as is greatly impacts the timescale on which fragmentation operates along with the size distribution of the fragments generated. This may be especially important when considering rubble pile like objects \cite{Krivov_2017} that may be present in the disk. Furthermore the treatment of the inner disk edge warrants further investigation as we expect many of the current model assumptions to not hold up in this environment.

\begin{acknowledgements} We acknowledge the support from the Swiss National Science Foundation (SNSF) under grant 200020\textunderscore192038. We would like to thank the anonymous referee for the valuable comments and suggestions that helped us improve the manuscript.
\end{acknowledgements}

\appendix
\section{Simulation details}

\begin{table}[ht]
    \centering
    \caption{Additional parameters for the test comparison vs \cite{Ormel_2012}}
    \begin{tabular}{cc}
    \hline 
    \hline
     Parameter & Value \\
    \hline 
    \hline
    Stellar mass & 1 $M_\oplus$  \\
    Viscosity  $\alpha$ & $10^{-4}$ \\
    Initial Embryo Mass & $10^{-6}M_\oplus$ \\
    \end{tabular}
    \label{tab:param_test}
\end{table}

\begin{table}[ht]
\centering
\caption{Specific material strengths for Ices and basalt at different relative velocities}
\begin{tabular}{ccccc}
    \hline 
    \hline
    type of planetesimals & $Q_{0s}$ & $Q_{0g}$ & $b_s$ & $b_g$\\
    \hline
    Basalt $5 km/s$ & 9e7 & 0.5 & -0.36 & 1.36\\
    Basalt $3 km/s$ & 3.5e7 & 0.3 & -0.38 & 1.36\\
    Basalt $25 m/s$ & 1.23e7 & 6.3e-8 & -0.31 & 2.27\\
    \hline
    Ices $3 km/s$ & 1.6e7 & 1.2 & -0.39 & 1.26\\
    Ices $0.5 km/s$ & 7e7 & 2.1 & -0.45 & 1.19\\
    \hline
\end{tabular}
\label{tab:Qd}
\end{table}

\begin{table}[ht]
    \centering
    \caption{additional global parameters for all of the populations}
    \begin{tabular}{ccc}
    \hline
    \hline
    Parameter & Value \\
    \hline 
    Stellar mass & 1 $M_\oplus$  \\
    Viscosity  $\alpha$ & $2 \times10^{-3}$ \\
    Power-law Gas    & $-0.9$ \\
    Power-law Solids & $-1.5$\\
    Initial Embryo Mass & $10^{-2}M_\oplus$ \\
    Number of Embryos & $1,20$\\
    Formation Time &  $2 \times10^{7}$yr \\
    \hline
    \end{tabular}

    \label{tab:param_syn}
\end{table}

\begin{table}[ht]
    \centering
    \caption{initial conditions of system 63 of population $L$}
    \begin{tabular}{cc}
    \hline
    \hline
    Parameter & Value  \\
    \hline
    $r_{in}$ &  $8.07 \times10^{-2}$ \\
    core radius & $1.26\times10^{2}$ \\
    $\Sigma_g$ &  $1.98\times10^{2}$ \\
    $\dot{M}_{wind}$  & $6.20\times10^{-7}$ \\
    $f_{D/G}$ &  $1.35\times10^{-2}$ \\
    \end{tabular}
    \label{tab:NGPPS11_init}
\end{table}

\section{Stirring functions}\label{sec:stirr}
The stirring functions for the viscous stirring of planetesimals from \citep{Ohtsuki_2002} are given by
\begin{align}
    P_{VS} &= \frac{73 \Tilde{e}^2}{10\Lambda^2} \text{ln}(1+10\Lambda^2/\Tilde{e}^2) + \frac{72I_{PVS}(\beta)}{\pi \Tilde{e}\Tilde{i}}\text{ln}(1+\Lambda^2) \\
    Q_{VS} &= \frac{4\Tilde{i}^2+0.2\Tilde{i}\Tilde{e}^3}{10\Lambda^2\Tilde{e}}\text{ln}(1 + 10\Lambda^2\Tilde{e}) + \frac{72 I_{QVS}(\beta)}{\pi \Tilde{e}\Tilde{i}} \text{ln}(1+\Lambda^2)\\
 P_{DF} &= \frac{ \Tilde{e}^2}{\Lambda^2} \text{ln}(1 + 10\Lambda^2 ) + \frac{576I_{Pdf}(\beta)}{\pi \Tilde{e} \Tilde{i} } \text{ln}(1  + \Lambda^2) \\
    Q_{DF} &= \frac{\Tilde{i}^2}{\Lambda^2} \text{ln}(1 + 10\Lambda^2) + \frac{576I_{Qdf}(\beta)}{\pi \Tilde{e} \Tilde{i} } \text{ln}(1 + \Lambda^2) \\
    \Lambda & = 1/12 (\Tilde{e}^2 +\Tilde{i}^2)*\Tilde{i} .
\end{align}
The $\Tilde{e}$ and $\Tilde{i}$ refer to the reduced eccentricity and inclination given by: $\Tilde{e} = (e_i^2+e_j^2)/h_m$ where $h_m= \big(\frac{m_i +m_j}{3M_\odot}\big)^{1/3}$ is the mutual hill radius of swarms of solids $i$ and $j$. Since it is impractical to calculate the elliptic integrals appearing in the in the equations above we need approximations. The integrals $I_{[P,Q],[VS,DF]}$ are approximated in the range $0<\beta = i/e <1$ as 
\begin{align}
    I_{PVS} &= \frac{\beta - 0.36251}{ 0.061547 + 0.16112\beta + 0.054473\beta^2}\\
    I_{QVS} &= \frac{0.71946 - \beta}{ 0.21239 + 0.49764\beta + 0.14369\beta^2}\\
    I_{PDF} &= \frac{98.912 + 38.384\beta + 0.209\beta^2}{51.996 + 127.503\beta + 49.781\beta^2} \\
    I_{QDF} &= \frac{ -9.562\cdot 10^{-4} +179.7\beta + 12.083\beta^2}{228.8 + 570.4\beta+ 234.1\beta^2} .
\end{align}

The approximations of $I_{PVS}$ and $I_{QVS}$ are given by \cite{Chambers_2006} and $I_{PDF}$ and $I_{QDF}$ are obtained in the same way and it can be checked that they match the approximated integrals from \citep{Ohtsuki_2002} within $3\%$ in the range $0<\beta\leq 1$ which is the range of allowed values in the code.

\bibliographystyle{bibtex/aa}
\bibliography{bibtex/references.bib}

\end{document}